\documentclass[prb,aps,twocolumn,footinbib,longbibliography,superscriptaddress,times,floatfix]{revtex4-2}

\usepackage{CJK}
\usepackage[colorlinks,bookmarks=false,citecolor=blue,linkcolor=red,urlcolor=blue]{hyperref}
\usepackage{color} 
\usepackage{graphicx}
\usepackage{float}
\usepackage{multirow}
\usepackage{amsmath}
\usepackage{amsfonts}
\usepackage{bm}
\usepackage[dvipsnames]{xcolor}
\usepackage[normalem]{ulem}

\usepackage[caption=false,subrefformat=parens,labelformat=parens]{subfig}

\newcommand{\CCO}{Ca$_{10}$Cr$_7$O$_{28}$}

\makeindex
\begin{document}
\graphicspath{{./figures/}}
\begin{CJK*}{UTF8}{gbsn} 

\title{Pinch points and half moons encode Berry curvature}



\author{Han Yan (闫寒)}
\affiliation{Theory of Quantum Matter Unit, Okinawa Institute of Science and Technology Graduate University, Onna-son, 
	Okinawa 904-0412, Japan}
\affiliation{Department of Physics \& Astronomy, Rice University, Houston, TX 77005, USA}
\affiliation{Smalley-Curl Institute, Rice University, Houston, TX 77005, USA}

\author{Judit Romh\'anyi}
\affiliation{Theory of Quantum Matter Unit, Okinawa Institute of Science and Technology Graduate University, Onna-son, 
	Okinawa 904-0412, Japan}
\affiliation{Department of Physics and Astronomy, University of California, Irvine, California 92697, USA}

\author{Andreas Thomasen}
\affiliation{Theory of Quantum Matter Unit, Okinawa Institute of Science and Technology Graduate University, Onna-son, 
	Okinawa 904-0412, Japan}
	
\author{Nic Shannon}
\affiliation{Theory of Quantum Matter Unit, Okinawa Institute of Science and Technology Graduate University, Onna-son, 
Okinawa 904-0412, Japan}

\date{\today}

\begin{abstract}

``Half moons'', distinctive crescent patterns in the dynamical structure factor, have been 
identified in inelastic neutron scattering experiments for a wide range of frustrated magnets.
In an earlier paper [H. Yan {\it et al.}, Phys. Rev. B~{\bf 98}, 140402(R) (2018)] we have 
shown how these features are linked to the local constraints realized in classical spin liquids.
Here we explore their implication for the topology of magnon bands. 
The presence of 
half moons indicates a separation of magnetic degrees 
of freedom into irrotational and incompressible components.
Where bands satisfying these constraints meet, it is at a singular point encoding Berry curvature of $\pm 2\pi$.
Interactions which mix the bands open a gap, resolving the 
singularity, and leading to bands with finite Berry curvature, accompanied by 
characteristic changes to half--moon motifs.   
These results imply that inelastic neutron scattering can, in some cases, be used 
to make rigorous inference about the topological nature of magnon bands.

\end{abstract}
\maketitle
\end{CJK*}
\section{Introduction}

The realisation that bands of electrons can be classified using topological indices 
provided both a monumental shock, and an enormous stimulus to research 
in condensed matter physics.
First studied as ``Chern Insulators'', in the context of the Integer Quantum 
Hall effect \cite{Klitzing1980,Laughlin1981,Thouless1982,Haldane1988}, and later 
revived in the context of Graphene \cite{Kane2005,Kane2005b}, 
work on topological bands of electrons has grown to encompass a large and active field 
of research on topological semi--metals, insulators and superconductors \cite{Hasan2010,Qi2011}.
More recently, magnetic insulators have also entered the stage, through 
the realization that bands of magnetic excitations can also carry 
topological indices.
A prime example is provided by the ``Shastry--Sutherland'' magnet  
SrCu$_2$(BO$_3$)$_2$ \cite{Kageyama1999}, where Dzyaloshinski--Moriya  
interactions both enable triplon excitations to form a dispersing band, and 
 act as a pseudo--magnetic field for these excitations, endowing 
them with a finite Chern number \cite{Romhanyi2015}.


Work on topological bands in magnets has evolved into an active subfield 
in its own right 
\cite{Shindou2013,Mook2014-PRB89,Mook2014-PRB90,Romhanyi2015,Owerre2016,Chernyshev2016,Zyuzin2016,Kim2016,McClarty2017,Nakata2017,McClarty2018,Kondo2019-PRB99,Joshi2019,Romhanyi2019,McClarty2019,Kondo2019-PRB100,Kawano2019,Mook2019,Malki2020,Bhowmick2020,Aguilera2020,Kondo2020,Mook2021,Kondo2021,Zhuo2021,Thomasen2021,Akagi2022,Neumann2022,Xing2022,Chen2022,Zhuo2022,Fujiwara2022,McClarty2022}, with important 
themes including the taxonomy of bosonic Chern insulators \cite{Kondo2019-PRB99,Thomasen2021}, the analysis of interactions 
\cite{Chernyshev2016,McClarty2018,McClarty2019}, and closing the gap between theory and 
experiment \cite{Katsura2010,Onose2010,Matsumoto2011,Ideue2012,Matsumoto2014,Hirschberger2015,McClarty2017,Cai2021,Zhang2021,Scheie2022,Suetsugu2022}.
This last point, however, presents a serious challenge.     
In the case of topological insulators, both the quantization of the Hall response,  
and the character of surface states, provide information about the topology 
of the underlying electron bands, with surface states easily accessible through 
tunneling or photoemission experiments \cite{Hasan2010,Qi2011}.
Equivalent surface states exist in topological bands of magnons \cite{Mook2014-PRB90}
and triplons \cite{Romhanyi2015}, but at present there is no established technique 
for measuring them.
Meanwhile, the corresponding thermal Hall and Nernst effects, while sensitive 
to Berry phases, are not quantized \cite{Katsura2010,Matsumoto2011,Matsumoto2014}, 
and notoriously difficult to measure.
It is therefore of great interest to know whether topological bands of excitations 
in magnetic insulators have other, more--easily accessible fingerprints in experiment.
 

\begin{figure}[t]
	\centering
	\subfloat[Topologically critical\label{fig:summary.trivial}]{
	\includegraphics[width=0.45\columnwidth]{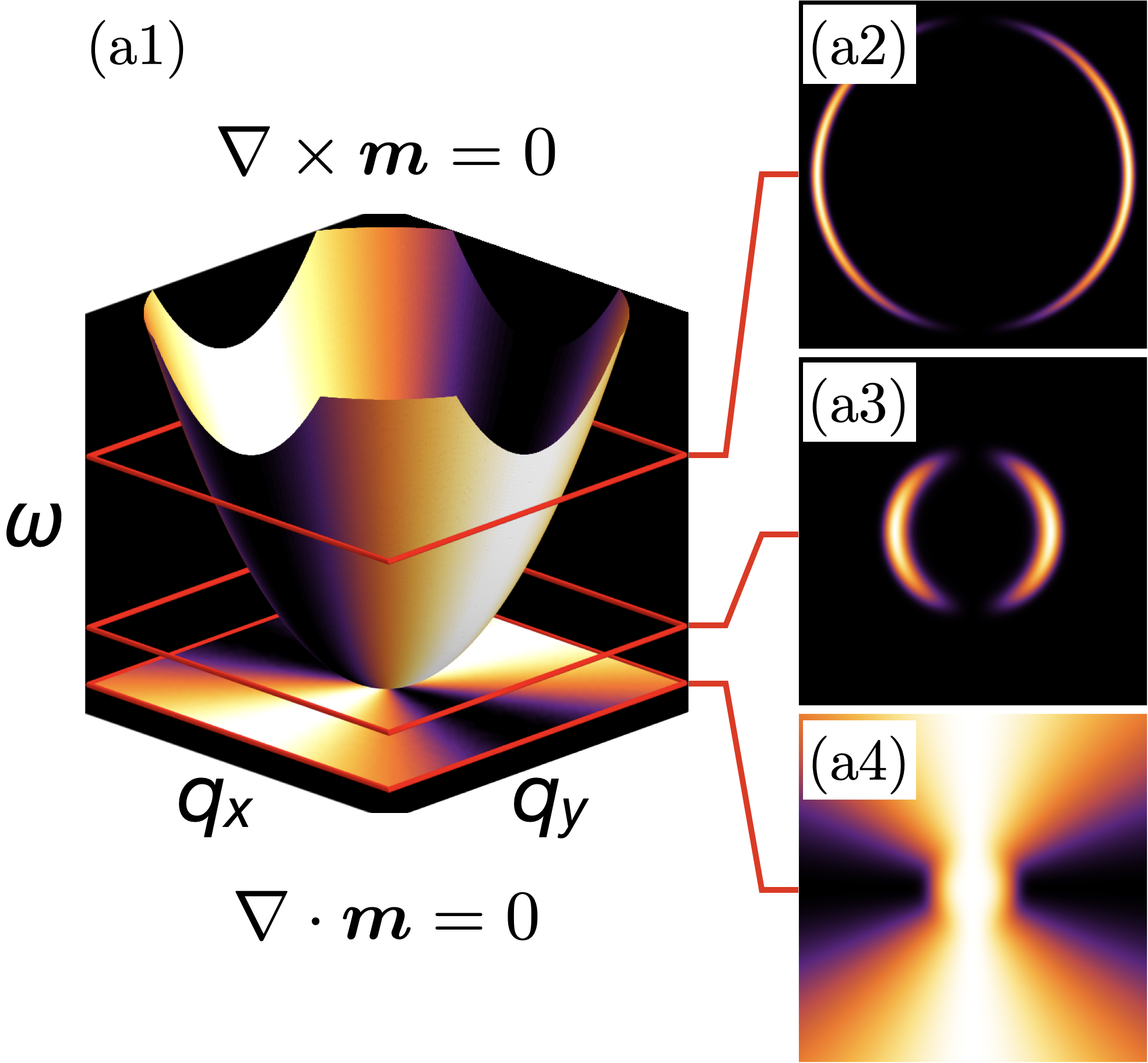}} 
	\subfloat[ Bands with Berry curvature\label{fig:summary.topological}]{
	\includegraphics[width=0.45\columnwidth]{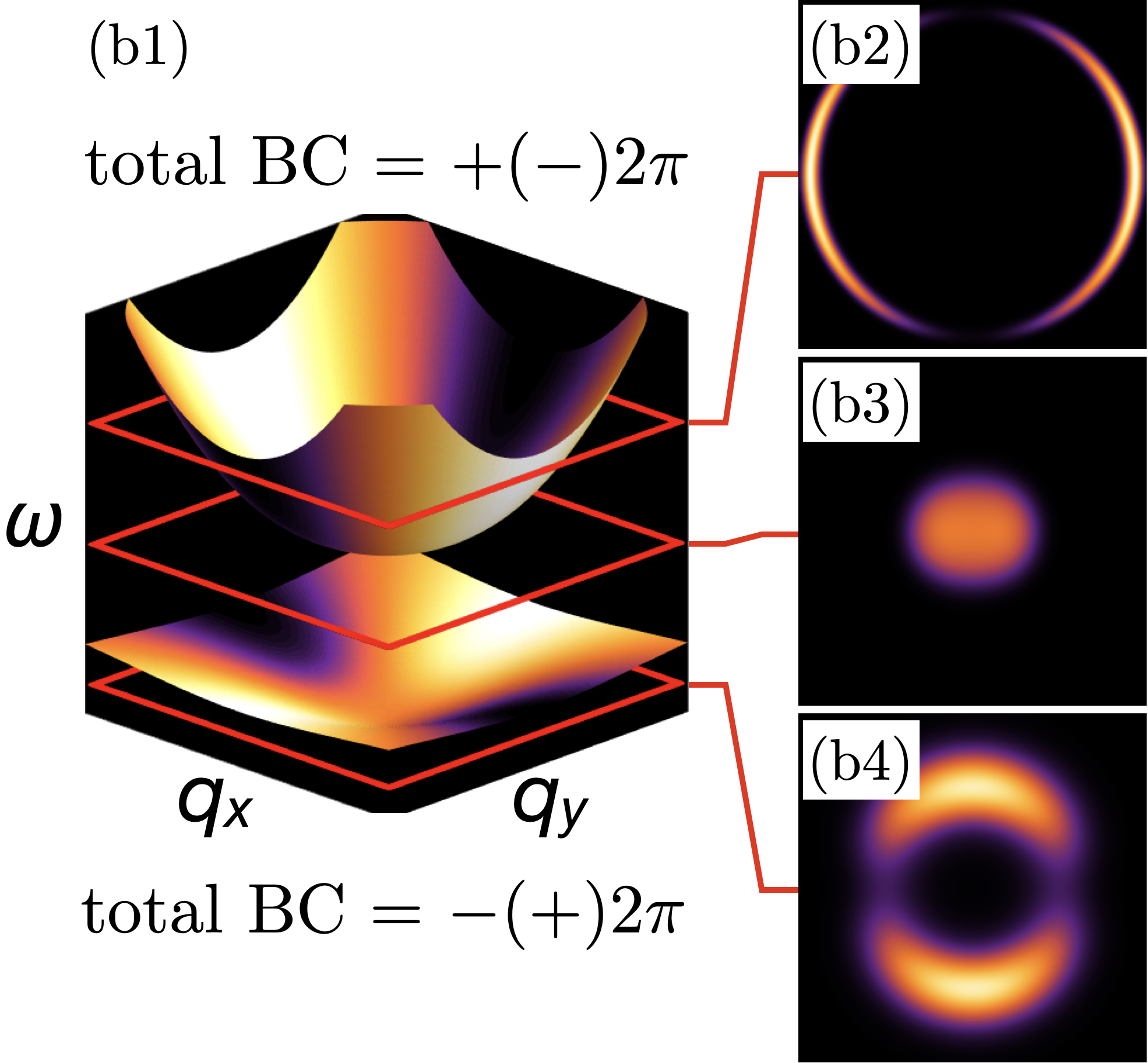}} \\
	\caption{
	Systematic modification of pinch--point and half--moon features in a 
	system with topological bands.
	a) Case with topologically critical bands, considered in \cite{Yan2018}.  
	Pinch points are inscribed on both flat and dispersing bands, 
	which meet in the zone center.
	Cuts through the dispersing band at fixed energy reveal 
	crescent--shaped half--moon features.
	The band--touching point is singular, and has a localized Berry 
	curvature (BC) of $\pm 2\pi$.
	b) The introduction of interactions which mix 
	states in the two bands opens a gap, eliminating the singular correlations 
	at the band--touching point, and endowing each band with BC of $\pm 2\pi$.
	       	}
	\label{fig:summary}
\end{figure}


In this work, we explore signatures of band--topology which are accesible in 
conventional, bulk, inelastic neutron scattering, in a broad class of frustrated magnets.
Building on earlier work  \cite{Yan2018}, we investigate 
the interplay between the widely observed  ``pinch--point'' and ``half--moon'' motifs,  
and the Berry curvature of associated band--excitations.
Pinch--points and half--moons arise as a consequence of ``fragmentation'' 
--- the separation of magnetic degrees of freedom into components with 
divergence-free (incompressible) and curl-free (irrotational) character \cite{Brooks-Bartlett2014,Yan2018,Mizoguchi2018}.
Where bands with divergence--free and curl--free character meet, 
pinch--point and half--moon features converge on a singular point 
with incipient Berry curvature of $\pm 2 \pi$ [Fig.~\ref{fig:summary.trivial}].
If interactions mix states belonging to these two bands, a gap opens, 
eliminating the singularity, and endowing each band with a Berry curvature $\pm 2 \pi$. 
This is accompanied by characteristic changes in pinch--point and 
half--moon features approaching the avoided band--touching 
[Fig.~\ref{fig:summary.topological}].


We demonstrate these effects for magnon bands in a model of spin--1/2 moments 
on the Kagome--lattice, where anisotropic exchange interactions provide the 
driving force for band topology.
None the less, the resulting phenomenology is very general, and we discuss a range   
of kagome and pyrochlore systems where pinch points and half moons 
herald a topological band structure, as well as developing detailed  
predictions for the Kagome ferromagnet Cu(1,3-bdc) [\onlinecite{Chisnell2015}].  
These results provide an explicit connection between the problem of a singular touching 
between flat and quadratically dispersing bands, 
 previously studied in the context of topological
 electron bands \cite{Chong2008,Bergman2008,Sun2009,Dora2014,Montambaux2018,Rhim2019,Hwang2021,Rhim2021},  
and spectral features routinely measured in frustrated magnets. 
  
\section{The Model}
\subsection{Kagome model and its magnon bands}
\begin{figure}[t]
	\centering
	\includegraphics[height=3cm]{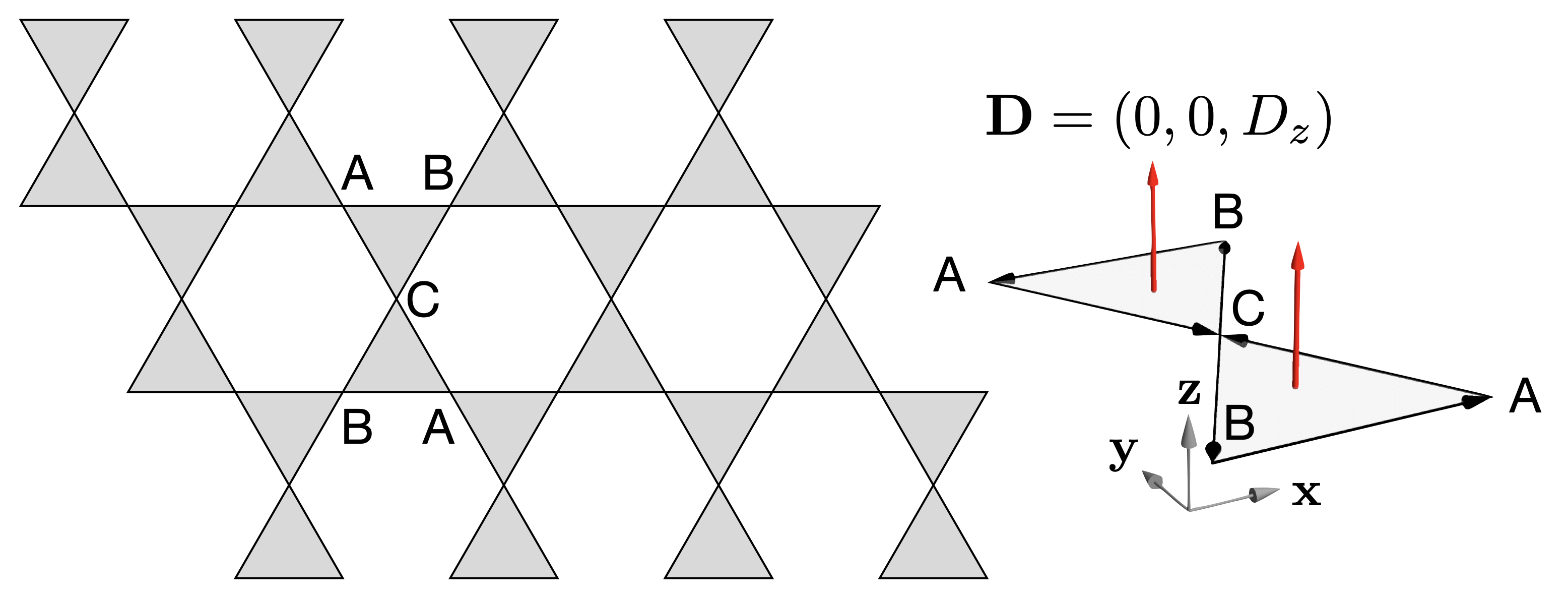}
	\caption{Kagome lattice, showing geometry of corner--sharing triangles, 
	and convention for labeling sites (A,B,C) within 3--site primitive unit cell.   
	Inset: Black circulating arrows
	indicate the sense in which  
	bonds are counted for Dzyaloshinskii--Moriya (DM) interactions.}
	\label{fig:lattice}
\end{figure}

In magnets, as in systems with itinerant electrons, 
topological bands usually originate in spin--orbit coupling.
This commonly takes the form of Dzyaloshinski--Moriya (DM) interaction 
with many of the most widely--studied frustrated lattices having a corner--sharing geometry which permits anisotropic 
exchange, including DM interactions \cite{Mook2014-PRB89,Mook2014-PRB90,Romhanyi2015,Chernyshev2015,Owerre2016,Chernyshev2016,Maksimov2016,Zyuzin2016,Kim2016,McClarty2017,Nakata2017,Kondo2019-PRB99,Joshi2019,Romhanyi2019,McClarty2019}.
As a representative example, we consider a spin--1/2 magnet with anisotropic exchange 
interactions on the first--neighbor bonds of a kagome lattice [Fig.~\ref{fig:lattice}].
The point--group symmetry of this lattice is D$_{6h}$ \cite{Essafi2017,Thomasen2021-Thesis}, 
and the existence of a mirror plane restricts anisotropic interactions to transverse spin components, 
so that the most general model is
\begin{eqnarray}
\mathcal{H} &=&
J_\perp \sum_{\langle ij \rangle} (S^x_i S^x_j +S^y_i S^y_j)+J_z \sum_{\langle ij \rangle} S^z_i S^z_j 
 	\nonumber\\
 && + D_z \sum_{\langle ij \rangle} (\bm{S}_i\times\bm{S}_j)_z + K_\perp \sum_{\langle ij \rangle} \mathbf{n}_{ij} \cdot \mathbf{Q}^{\perp}_{ij}
 \nonumber\\
 &&  - g_z h^z\sum_i S^z_i \; ,
\label{eq:H}
\end{eqnarray}
where 
$h^z$ is an applied magnetic field, $\mathbf{n}_{ij}$ is a unit vector in the direction 
of the bond $ij$, and 
\begin{eqnarray}
	\mathbf{Q}^{\perp}_{ij} &=& (S^x_i S^x_j - S^y_i S^y_j,\ S^x_i S^y_j + S^y_i S^x_j ) \; ,
\end{eqnarray}
\footnote{
This parameterisation of interactions is exactly equivalent to that given in \cite{Essafi2017,Thomasen2021-Thesis}, 
with $K_\perp$ in Eq.~(\ref{eq:H}) playing the role of $J_x - J_y$ in \cite{Essafi2017,Thomasen2021-Thesis}.}.
In what follows we will emphasize DM interactions $D_z$, 
setting $J_\perp=J_z=J$ and $K_\perp =0$.
None the less, the conclusions we reach about the relationship between ``half moons'' 
and Berry curvature are completely general, and hold regardless of which type of anisotropy 
is considered. 
A complete analysis, including $K_{\perp}$, is given in Supplemental Material \cite{supplemental-material} (see also references \cite{Essafi2017,Thomasen2021-Thesis,Balian1969,Thomasen2021,MacDonald1988,Yan2018} therein).


We focus on single--magnon excitations about a fully polarized state, stabilized by 
either ferromagnetic (FM) exchange interactions \mbox{$J < 0$}, or magnetic field \mbox{$h^z \gg J$}.
For sufficiently large spin--wave gap (guaranteed by high magnetic field), 
these will be well--described by a non--interacting theory \cite{Chernyshev2016,McClarty2018,McClarty2019}, 
which we can access through the equation of motion (EoM) 
$-i \hbar\partial_t a^\dagger_{\lambda, {\bf q}} = [{\mathcal H}, a^\dagger_{\lambda, {\bf q}}] $
for the magnon creation operator 
\begin{eqnarray}
	a^\dagger_{\lambda, {\bf q}} = \frac{1}{\sqrt{N}} \sum_{i} \phi_{\lambda, i} S_i^- e^{-{\bf q} \cdot {\bf r}_i,}  
	\; ,	\label{eq:one.magnon}
\end{eqnarray}
where \mbox{$\lambda= 1,2,3$} is the band index, and all information about band topology is encoded in $\phi_{\lambda, i}$.  


\begin{figure*}[htb]
	\subfloat[\label{Fig.Kagome.SW.dispersion} Band dispersion, \hspace{3.5cm} $D_z = 0$]{\includegraphics[height=4.5cm]{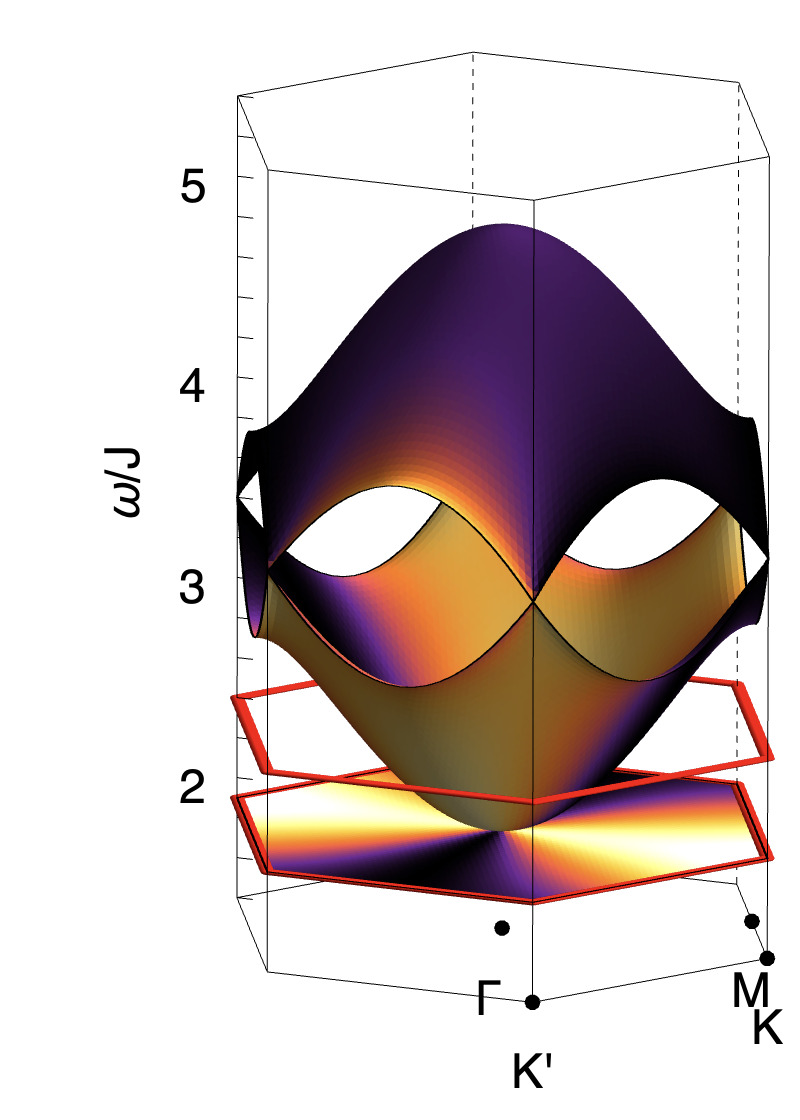}\includegraphics[height=2.5cm]{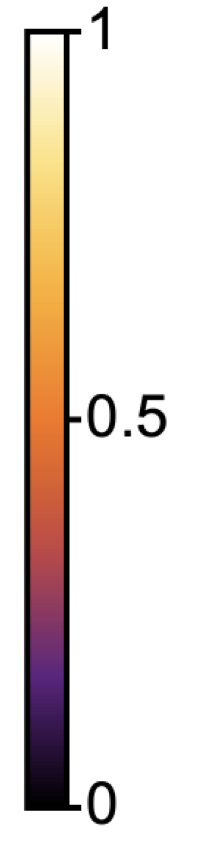} }
	\subfloat[\label{Fig:Berry.curvature.Dz.zero} Berry curvature, \hspace{1cm}$D_z = 0$]{\includegraphics[height=4.5cm]{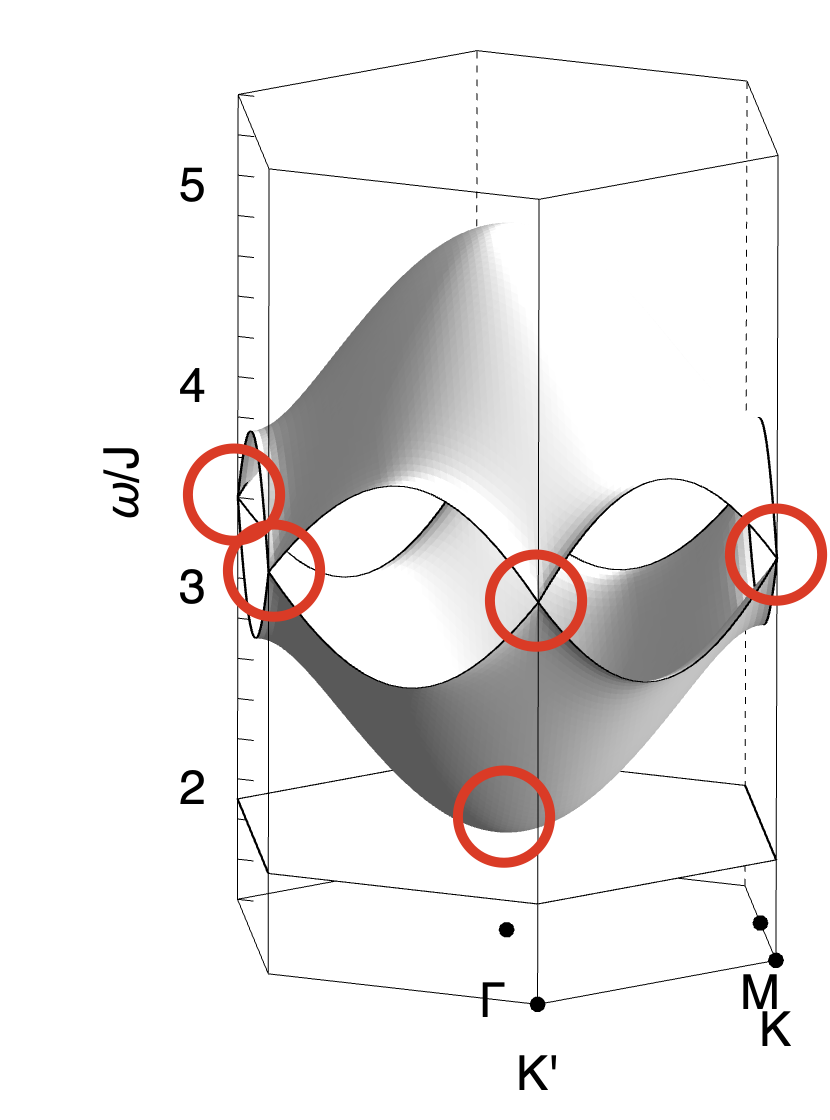} \includegraphics[height=2.5cm]{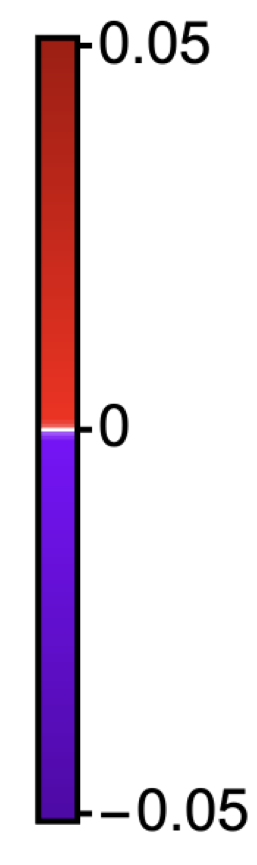} }
	\subfloat[\label{Fig.Kagome.SW.dispersion.with.D} Band dispersion, \hspace{3.5cm} $D_z = 0.1$]{\includegraphics[height=4.5cm]{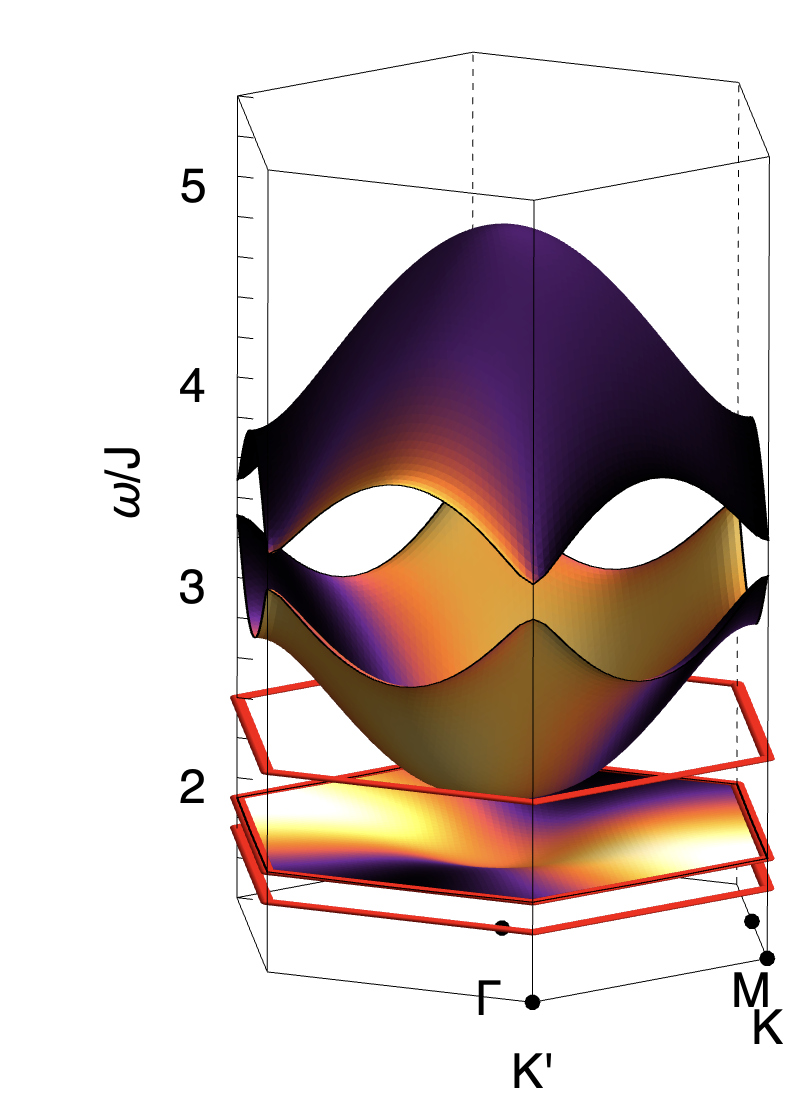} }
	\subfloat[\label{Fig:Berry.curvature.Dz.finite} Berry curvature, \hspace{3.5cm} $D_z = 0.1$]{\includegraphics[height=4.5cm]{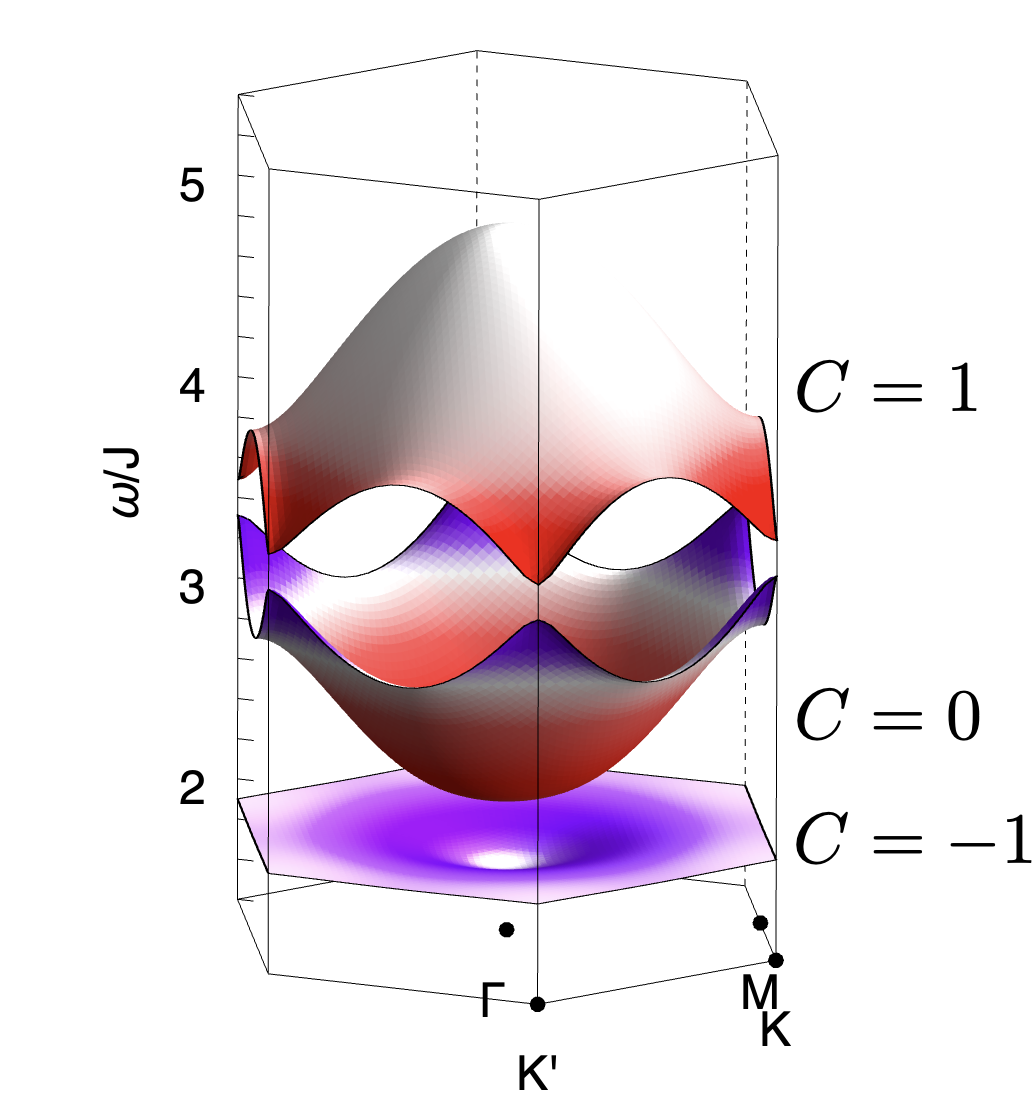} } \\
	\subfloat[\label{Fig.Kagome.SW.1} $S({\bf q}, \omega = 2.0)$, \hspace{3.5cm}  $D_z = 0$]{\includegraphics[height=0.17\textwidth]{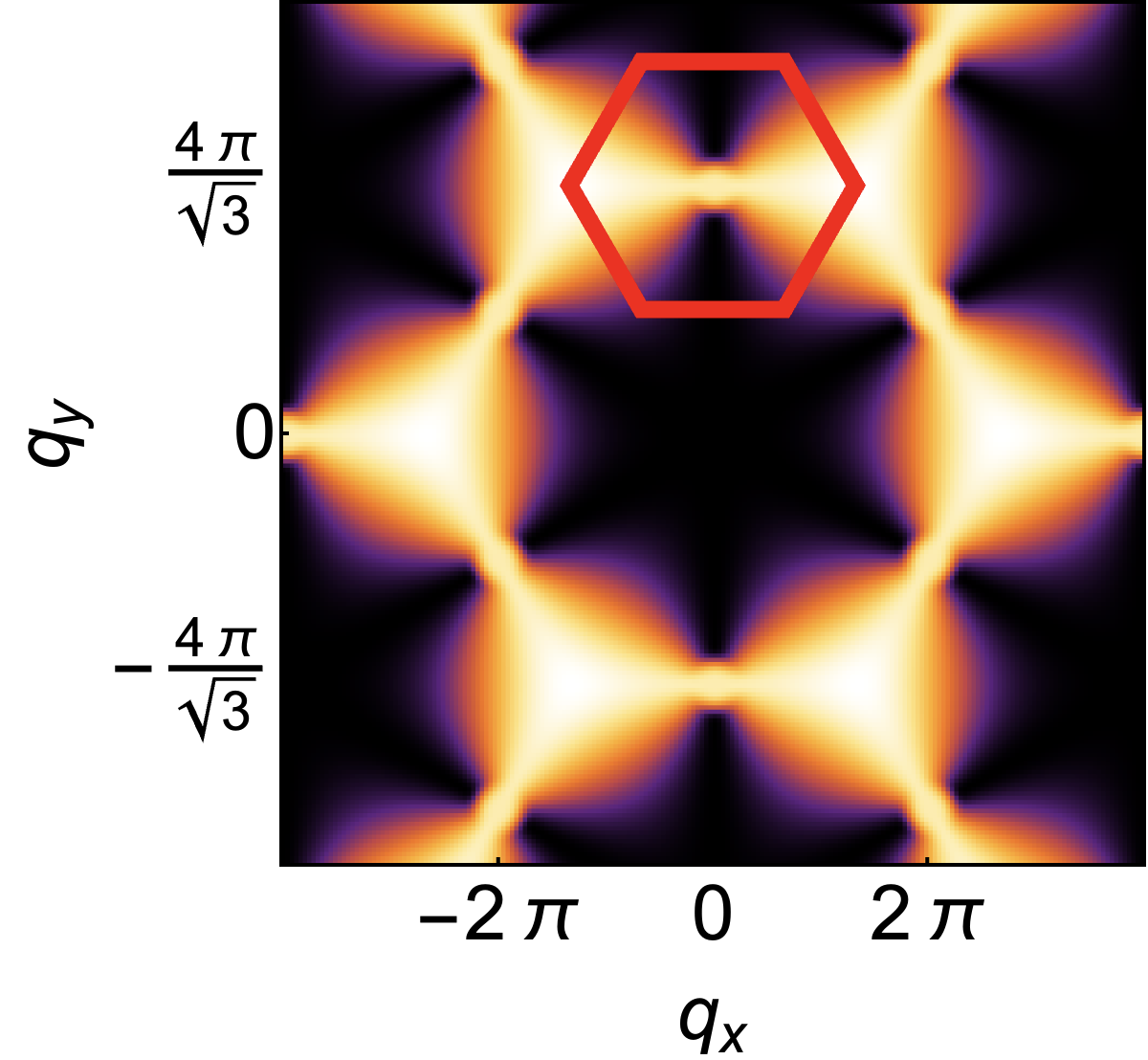}}
	\subfloat[\label{Fig.Kagome.SW.2} $S({\bf q}, \omega=2.5)$,  \hspace{3.5cm} $D_z = 0$]{\includegraphics[height=0.17\textwidth]{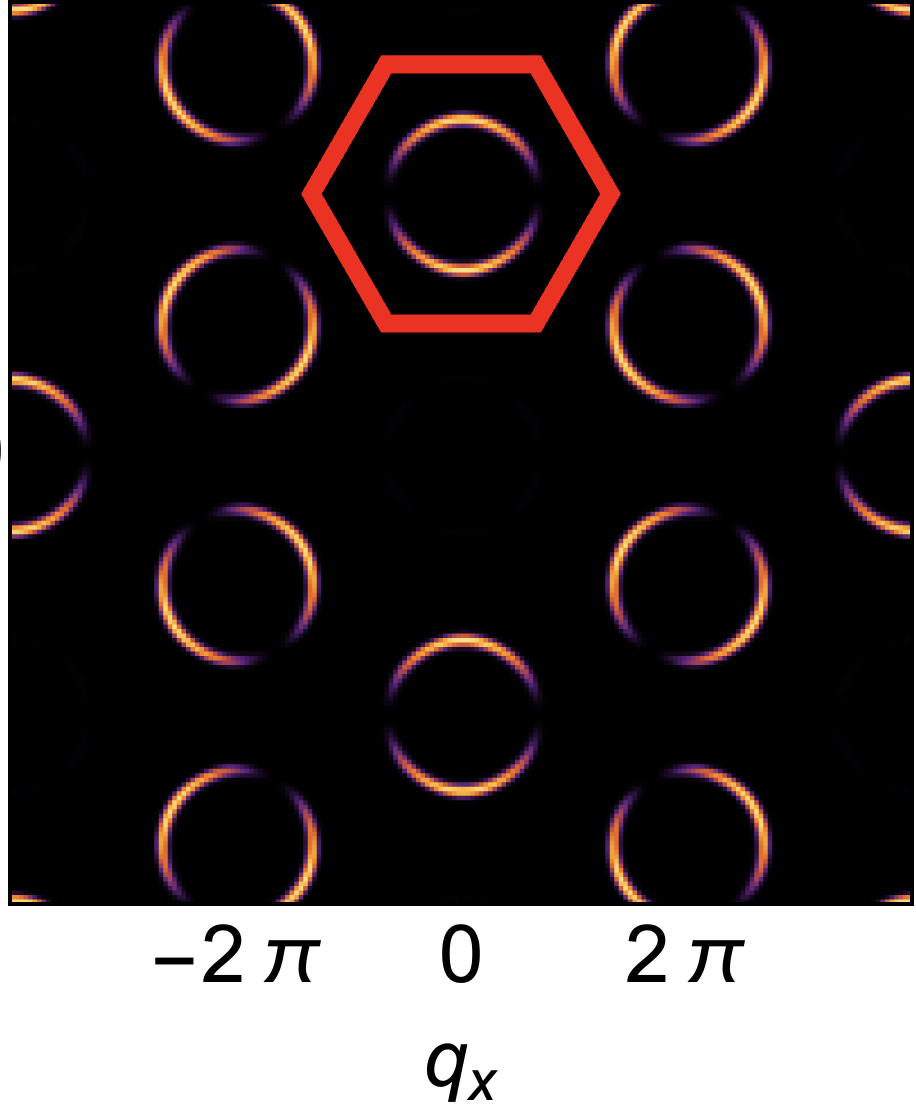}}\qquad 
	\subfloat[\label{Fig:LSW.Sqomega.withD.g} $S({\bf q}, \omega=1.85)$, \hspace{3.5cm}  $D_z = 0.1$]{\includegraphics[height=0.17\textwidth]{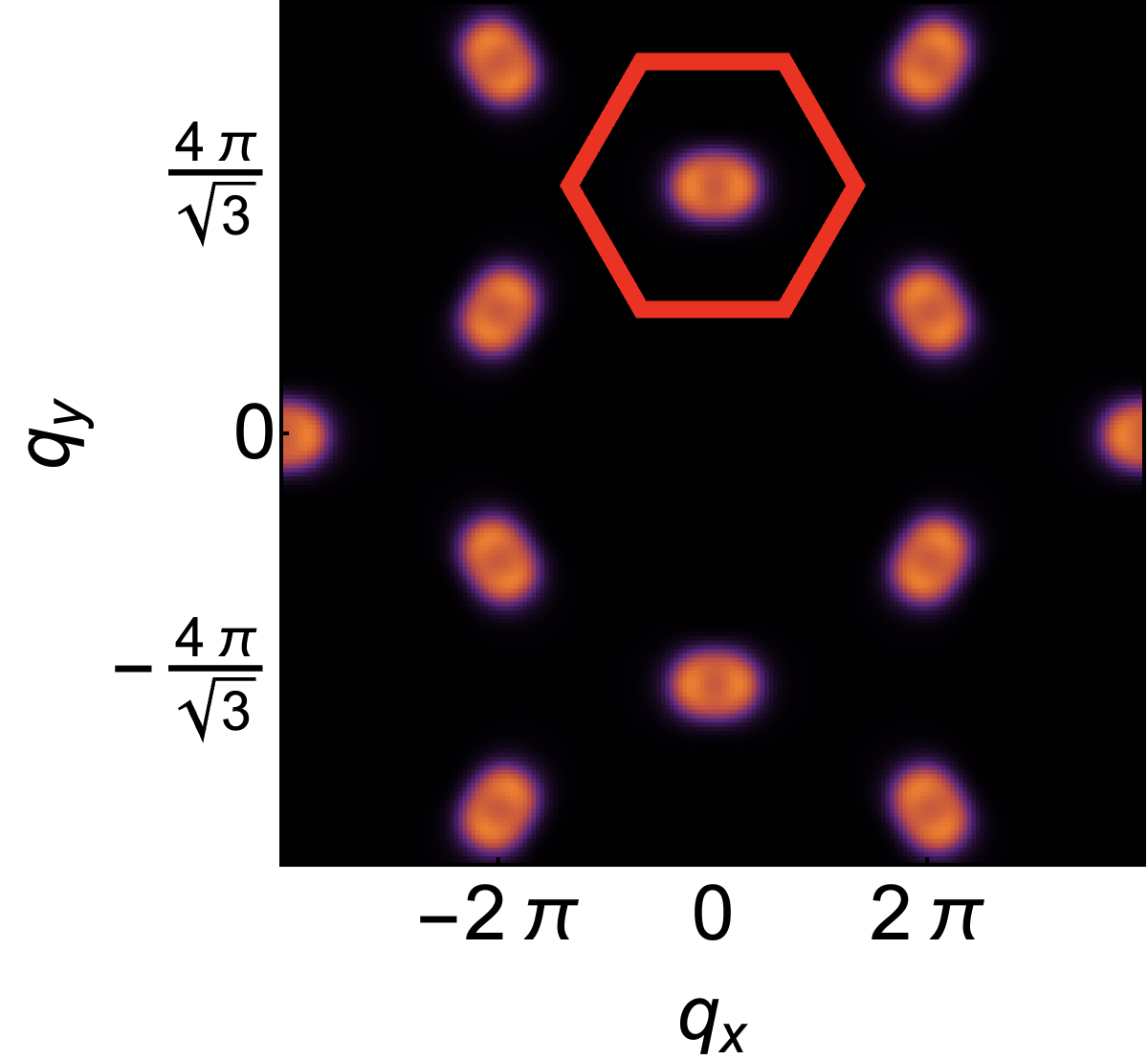} }
	\subfloat[\label{Fig:LSW.Sqomega.withD.h} $S({\bf q}, \omega=2.0)$, \hspace{3.5cm}  $D_z = 0.1$]{\includegraphics[height=0.17\textwidth]{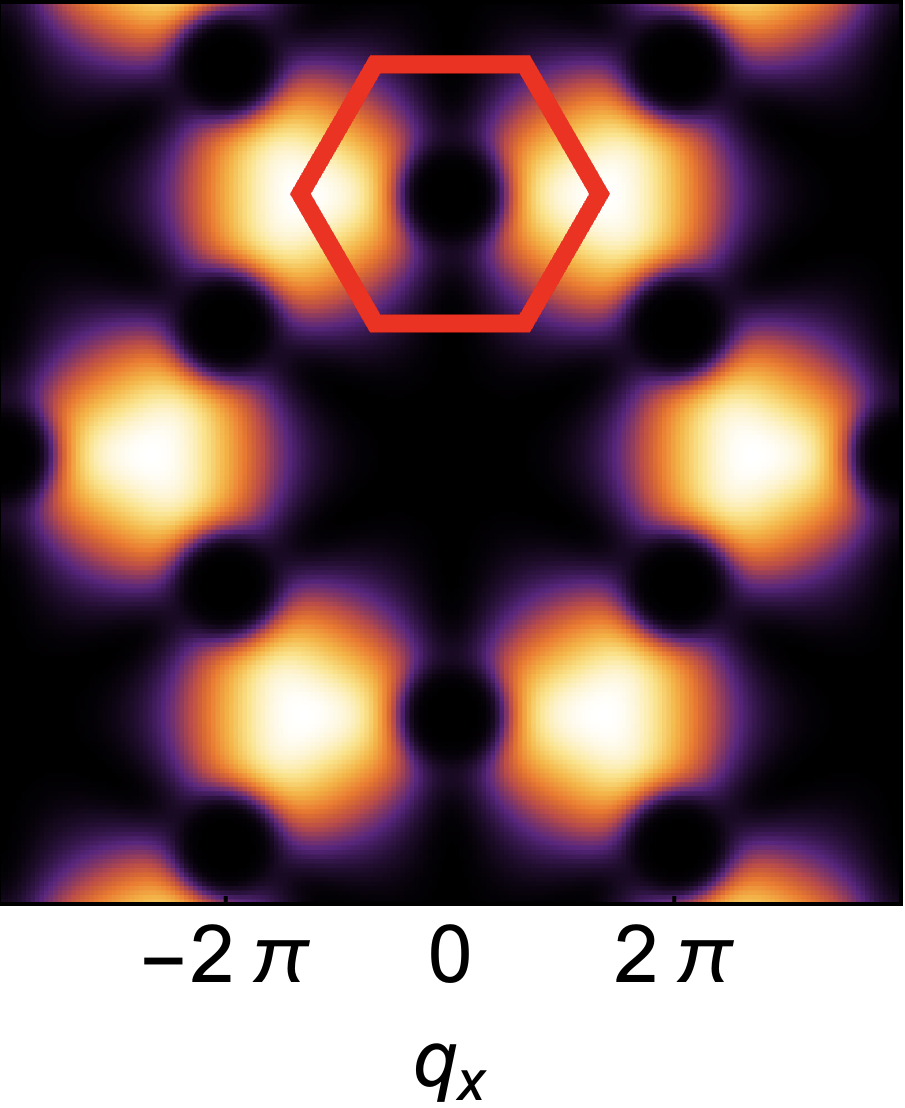} }
	\subfloat[\label{Fig:LSW.Sqomega.withD.i} $S({\bf q}, \omega=2.5)$, \hspace{3.5cm}  $D_z = 0.1$]{\includegraphics[height=0.17\textwidth]{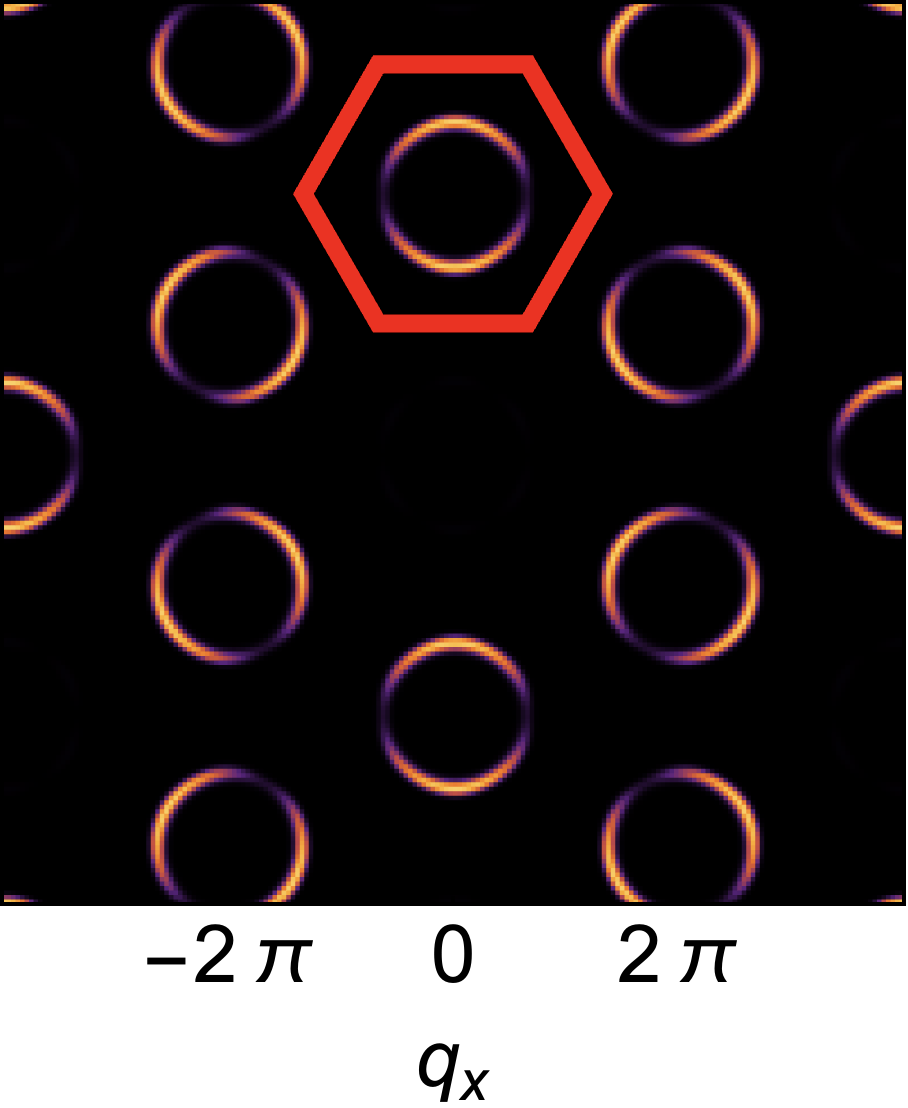} }
	\caption{
		Relationship between singular features in the dynamical structure factor $S({\bf q}, \omega)$, 
		and Berry curvature in Kagome lattice magnets with and without Dzyaloshinskii--Moriya (DM) interactions.
		(a)~Spin--wave dispersion of Heisenberg antiferromagnet 
		(HAF) on a Kagome lattice in high magnetic field.
		The colorscale shows how each band contributes to $S({\bf q}, \omega)$.
		Pinch points, singular for ${\bf q} \to \Gamma$, are inscribed on the flat band 
		at $\omega = 2$, and the dispersing band which touches it.
		(b)~Berry curvature associated with spin--wave bands.
		Curvature is localised at the topologically--critical band--touching points, shown with red circles, 
		and is zero elsewhere.
		(c)~Dispersion in the presence of finite DM interaction, $D_z =0.1$, 
		showing how the mixing of states between bands opens gaps at band--touching points.
		(d)~Berry curvature generated by DM interaction, through mixing of states.
		Integrated accross the Brillouin zone (BZ), this leads to bands with the Chern numbers $C = 1,~0,~-1$. 
		(e)~Dynamical structure factor $S({\bf q}, \omega)$ at $\omega = 2.0$, for $D_z =0$, 
		showing pinch points inscribed on flat band.
		Red hexagon denotes the BZ considered in (a)--(d).
		(f)~Equivalent results for $\omega = 2.5$, showing half moons associated with dispersing band.
		[cf. Fig.~\ref{fig:summary.trivial}].
		(g)~Structure factor for $\omega = 1.85$, $D_z =0.1$, 
		showing how the mixing of states between bands eliminates the pinch--point singularity for 
		${\bf q} \to \Gamma$.
		(h)~Equivalent results for $\omega = 2.0$, showing elimination of 
		singular features on the dispersing band.
		(i)~Equivalent results for $\omega = 2.5$, showing survival of half moons away from zone center.
		[cf. Fig.~\ref{fig:summary.topological}].
		All results were obtained within linear spin wave (LSW) theory for 
		Eq.~\eqref{eq:H}, for parameters $J  = 1$, $g_zh^z = 5$.
		Results in (e)--(i) have been convoluted with a Gaussian of FHWM = $0.1\ J$ 
		to mimic experimental resolution.}
	\label{fig:HAF.LSW}
\end{figure*}


\subsection{Symmetry analysis and Helmholtz decomposition}  
In order to make connection with singular features in scattering (half moons), 
we now decompose magnetic excitations  $\phi_{\lambda, i}$ into incompressible
(divergence--free) and irrotational (curl--free) components.
These correspond to
irreducible representations (irreps) 
of spins on the triangular units which form the building--blocks of the 
Kagome lattice
\begin{subequations}
\begin{eqnarray}
\Phi_{{\sf A}_1} &=&\frac{S^-_A+ S^-_B+ S^-_C}{\sqrt{3}} \;, 
\label{eq:A1.irrep} \\
{\bm m}_{\sf E} &=& 
	\left(\! \frac{S^-_A -S^-_B}{\sqrt{2}},\frac{S^-_A  +  S^-_B  - 2 S^-_C}{\sqrt{6}}\!\right) \;,
	\label{eq:E.irrep}
\end{eqnarray}
\label{eq:irreps}
\end{subequations}
where the convention for labelling lattice sites is given in Fig.~\ref{fig:lattice}.
In the long--wavelength limit (${\bm q} \to  {\bm 0} \equiv \Gamma $), this 
allows us to describe magnons,  Eq~(\ref{eq:one.magnon}), in terms of fields 
${\bm m}_{\sf E}$ and $\Phi_{{\sf A}_1}$.   
We then employ the Helmholtz--Hodge decomposition~\cite{Arfken1995}, writing 
\begin{eqnarray}
{\bm m}_{\sf E}   &=& {\bm m}_{\sf E} ^\text{\sf curl} + {\bm m}_{\sf E} ^\text{\sf div}  \; ,
\label{eq:helmholtz}
\end{eqnarray}
where 
$\nabla \cdot {\bm m}_{\sf E} ^\text{\sf curl} = 0$, 
$\nabla_\perp \cdot {\bm m}_{\sf E} ^\text{\sf div} = 0$, 
and $\nabla_\perp = (-\partial_y, \partial_x)$ is the two-dimensional curl.
From the Heisenberg EoM  
$-i\partial_t {\bm m}_{\sf E} = [{\mathcal H}, {\bm m}_{\sf E}]$
we find
\begin{subequations}
\begin{align}
	-i\partial_t {\bm m}_{\sf E}^\text{\sf curl}  & =  \omega_0 {\bm m}_{\sf E} ^\text{\sf curl} 
	- i  \sqrt{3}  D_z {\bm m}_{\sf E}^\text{\sf div}
	\; , \label{eq:EoM.incompressible}\\
	-i\partial_t {\bm m}_{\sf E}^\text{\sf div} & = - \rho_S \nabla (\nabla\cdot {\bm m}_{\sf E} ^\text{\sf div}) +  \omega_0 {\bm m}_{\sf E} ^\text{\sf div} 
	- i\sqrt{3}  D_z {\bm m}_{\sf E}^\text{\sf curl}
	\; ,\label{eq:EoM.irrotational}
\end{align}
\label{eq:EoM}
\end{subequations}
where $\omega_0=h - 3 J$,  $\rho_S =  \frac{1}{8} J$, and we set $\hbar = 1$.

\subsection{Origin of half moons}
In the absence of anisotropic exchange, i.e. for \mbox{$D_z=0$}, excitations 
with incompressible and irrotational character are decoupled \cite{Yan2018,Benton2016}.  
Solving Eq.~(\ref{eq:EoM.incompressible}) by Fourier transformation $-i\partial_\alpha\ \rightarrow \ q_\alpha$, 
we find a flat band of incompressible excitations 
\begin{eqnarray}
	\omega^\text{\sf curl} ({\bm q}) = \omega_0 \; , 	\label{eq:omega.curl}
\end{eqnarray}
with corresponding eigenvector 
\begin{eqnarray}
	\tilde{\bm m}_{\sf E} ^\text{\sf curl}({\bm q}) = (-q_y, q_x)/q  \label{eqn.m.curl.eigen} \; .
\label{eq:zero.divergence.eigenvector}
\end{eqnarray}


Meanwhile, making use of the identity $\nabla (\nabla\cdot {\bm m}_{\sf E}^\text{\sf div} ) 
=\nabla^{T} \nabla {\bm m}_{\sf E}^\text{\sf div}$, 
we can solve Eq.~(\ref{eq:EoM.irrotational}) to find a quadratically--dispersing 
band of irrotational excitations 
\begin{eqnarray}
	\omega^\text{\sf div} ({\bm q}) = \omega_0 + \rho_S q^2  \;,  \label{eq:omega.div}
\end{eqnarray}
with eigenvector 
\begin{align}
\tilde{\bm m}_{\sf E}^\text{\sf div}({\bm q}) = (q_x, q_y)/q  \label{eqn.m.div.eigen}  \; .
\end{align}
At ${\bm q}  =  {\bm 0}$ the flat and dispersing bands become degenerate, 
touching quadratically.
This band touching is accompanied by a singularity in the associated 
Bloch functions [Eq.~(\ref{eq:zero.divergence.eigenvector})], reflecting 
the topology of localized states  
\cite{Bergman2008,Rhim2019,Rhim2021,Hwang2021,Udagawa-arXiv}.


We are now in a position to calculate dynamical structure factors.
Writing ${\bm q} = q\ (\cos\theta,\sin\theta)$ we find 
\begin{eqnarray}
	S^{\alpha\beta}_ {\nu} ({\bm q}, \omega) 
	&\propto& 		
	\frac{1}{2} \left[ 
	\delta^{\alpha\beta}  
	\pm A^{\alpha\beta}(\theta)   
	f (q)
	\right]  
	\delta(\omega - \omega_\nu ({\bm q})) 
	\; , \nonumber\\
	\label{eq:S.q.omega}
\end{eqnarray}
where $\nu = \pm$ indexes the irrotational 
and incompressible modes, respectively;  
$\alpha, \beta = x, y$;   
\begin{eqnarray}
	A (\theta) 
	&=&  
	\begin{bmatrix}
		\cos(2\theta)  & \sin(2\theta)  \\
		\sin(2\theta) & -\cos(2\theta) 
	\end{bmatrix} \; ; 
\end{eqnarray}
and, for $D_z=0$, $f(q) \equiv 1$.


For $D_z=0$, the limit 
\mbox{$S^{\alpha\beta}_ {\nu} ({\bm q} \to {\bf 0}, \omega)$}  
is a function of $\theta$, and therefore singular.
This form of singularity is well--known in the context of spin--liquids, 
where it is referred to as a ``pinch--point'' \cite{Henley2010,Yan2018}.
For the flat band associated with incompressible excitations, 
the pinch point is directly visible in the dynamical structure factor at  
$\omega = \omega_0$ \cite{Yan2018,Benton2016}.
Meanwhile, for the dispersing band of irrotational excitations, 
the pinch--point manifests as  ``half--moons'' 
on constant--energy cuts for $\omega > \omega_0$ \cite{Yan2018}.
These effects are illustrated schematically in  Fig.~\ref{fig:summary.trivial}.
\footnote{Here and in Fig.~\ref{fig:HAF.LSW}, 
singular behaviour for ${\bf q} \to {\bf 0}$ is obscured by the finite energy 
resolution used in making plots of \mbox{$S ({\bm q}, \omega)$}.}.


Exactly the same phenomenology is found in direct calculations 
from the microscopic model Eq.~(\ref{eq:H}).
In Fig.~\ref{fig:HAF.LSW} we present results of linear spin wave 
(LSW) theory, showing the flat band and dispersing bands found 
for $J=1$, $h=5$, $D_z = 0$ [Fig.~\ref{Fig.Kagome.SW.dispersion}].
The pinch point on the flat band is exhibited in Fig.~\ref{Fig.Kagome.SW.1}, 
while half moons associated with the dispersing band are shown in 
Fig.~\ref{Fig.Kagome.SW.2}.
Details of these calculations are given in the Supplemental Materials 
\footnote{
For completeness, we note that pinch points and half moons 
can also arise in the ground--state correlations of Kagome magnets, where they 
provide information about spin liquid states and their 
instabilites \cite{Mizoguchi2018,Kiese2023}.}.


\subsection{Opening of gap and elimination of singularity} 
The new feature which arises in the presence of anisotropic exchange 
interactions is the mixing of incompressible and irrotational excitations.
Solving Eq.~(\ref{eq:EoM.incompressible}), Eq.~(\ref{eq:EoM.irrotational}) 
for \mbox{$D_z \ne 0$}, we find that the quadratic 
band--touching at \mbox{${\bf q} = {\bf 0}$} is replaced by a gap 
\begin{eqnarray}
	\Delta = 2\sqrt{3} |D_z| \; , 
\end{eqnarray}
between bands with dispersion
\begin{eqnarray}
\omega_\pm (\bm{q}) &=& \omega_0 +  \frac{1}{2} \rho_S q^2  
	\pm  \frac{1}{2} \sqrt{   \Delta^2 +\rho_S^2 q^4}  \; .
\label{eq:omega.q.pm}
\end{eqnarray}
Hybridization also eliminates singular correlations for \mbox{${\bf q} \to {\bf 0}$}; 
the dynamical structure factor
is given by Eq.~(\ref{eq:S.q.omega}), but with 
\begin{eqnarray}
f (q) 
	&\to&  
	\begin{cases}
	 4\times \left(\frac{q}{q_0}\right)^2  \; , \; q \ll q_0 \\
	  1  \; , \; q \gg q_0
	  \end{cases}
\; ,
\end{eqnarray}
where $q_0 = \sqrt{|\Delta/\rho_S|} \sim \sqrt{|D_z/J|} $
determines the scale at which correlations cross over from a
pinch--point structure at large $q$, 
to a non--singular behaviour for \mbox{$q \to 0$}.


As a consequence the pinch--points and half--moons observed 
for \mbox{$D_z = 0$}  vanish for $q \ll q_0$.
Nonetheless, half--moon features remain visible for $q \gg q_0$, 
and can also be resolved on the formerly--flat band associated with 
incompressible excitations, as a consequence of its finite dispersion.
This phenomenology is illustrated schematically in Fig.~\ref{fig:summary.topological}, 
and for LSW calculations in Fig.~\ref{Fig.Kagome.SW.dispersion.with.D} 
and Fig.~\ref{Fig:LSW.Sqomega.withD.g}--\ref{Fig:LSW.Sqomega.withD.i}

\section{Topology and Pinch Points}
\subsection{Berry phase effects and band--topology}
The Helmholtz decomposition, Eq.~(\ref{eq:helmholtz}) also has 
important implications for band topology.
In the absence anisotropic exchange interactions, 
each of the eigenvectors Eq.~(\ref{eqn.m.curl.eigen})  
and Eq.~(\ref{eqn.m.div.eigen}) winds exactly 
once around the singular point, ${\bm q} = {\bm 0}$.
As a consequence, their quadratic band--touching 
is topologically critical, encoding Berry curvature of $\pm 2\pi$ at the singularity
\textcolor{red}{\cite{Chong2008,Sun2009,Dora2014,Montambaux2018,Rhim2019,Thomasen2021}}.
Introducing an interaction which mixes  the  incompressible and irrotational 
excitations, such as $D_z$, resolves the singularity at $\Gamma$, opens a gap, 
and injects Berry curvature of $\pm 2\pi$ into each band.  


We can explore these effects in more detail by noting that 
the EoM, Eq.~(\ref{eq:EoM}), describe an effective two--level system, 
familiar from earlier work on topological electrons~\cite{Chong2008,Sun2009,Dora2014,Montambaux2018,Rhim2019}
\begin{eqnarray}
\mathcal{H}_{\sf eff} ({\bm q})
	= \epsilon_0 \mathbb{I}_2 + \boldsymbol{d}({\bm q}) \cdot \boldsymbol{\sigma}\;,
\label{eq:Dirac}
\end{eqnarray}
where $\mathbb{I}_2$ is a $2\times2$ identity matrix, $\boldsymbol{\sigma}$ 
is a vector of Pauli matrices, $\epsilon_0=\omega_0+\frac{\rho_S}{2} q^2$, and 
\begin{equation}
\boldsymbol{d}({\bm q})=\left(\! \rho_S q_x q_y, -\sqrt{3}D_z, \rho_S\frac{q^2_x-q^2_y}{2} \!\right)
\label{eq:d_vector} \; .
\end{equation}
The Berry curvature associated with $\boldsymbol{d}({\bm q})$ is 
\begin{eqnarray}
\omega^\pm_{xy} ({\bm q})
&=& \mp \frac 1 2 \hat{\bm d}(\partial_{q_x}  \hat{\bm d}\times \partial_{q_y} \hat{\bm d}) 
= \mp 2 \frac{\Delta \rho_S^2 q^2}{(\Delta^2+\rho_S^2 q^4)^{3/2}} \; .  
\label{eq:d}
\end{eqnarray}
This curvature integrates to $\pm 2\pi$, and is concentrated 
in the vicinity of \mbox{${\bm q}={\bm 0}$}.
In the absence of other sources of Berry curvature, 
this will endow magnon bands with integer Chern number \mbox{$C = \pm 1$}. 
This phenomenology, linked to a quadratic band--touching, 
should be contrasted with the widely--studied case of 
linear band--crossings (Dirac cones), which encode 
Berry curvature $\pm \pi$ \cite{Bernevig2013-Book}.


It is also possible to calculate both Berry curvature and Chern numbers 
within LSW theory.
Results for \mbox{$D_z = 0.1$} are shown in Fig.~\ref{Fig:Berry.curvature.Dz.finite}.
Once the third band is taken into account, magnon bands also receive a net 
contribution of Berry phase $2 \times \pm \pi$ from the linear 
band--crossings at $K$ and $K'$ [Fig.~\ref{Fig:Berry.curvature.Dz.zero}].
For \mbox{$D_z > 0$}, \mbox{$K_\|=0$}, this leads to bands with 
Chern numbers \mbox{$C = -1, 0, 1$}.


\begin{figure*}[ht]
	\centering
	\subfloat[Topological magnon bands]{\includegraphics[height=0.4\columnwidth]{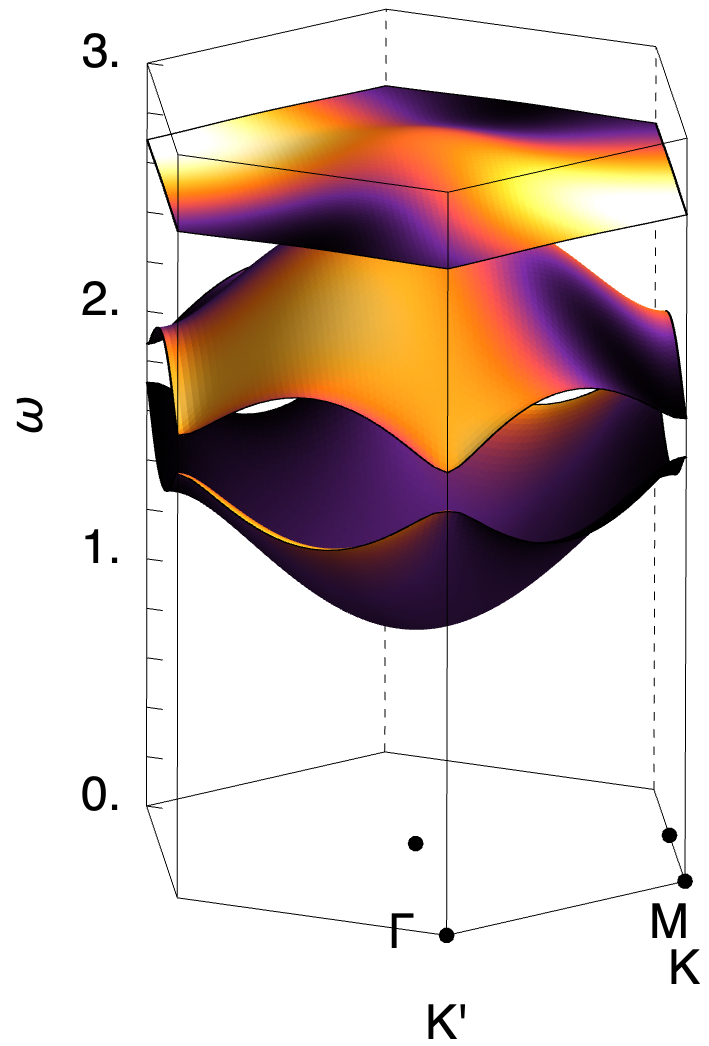}}\;
	\subfloat[Dynamical structure factor $S({\bm q},\omega)$]{\includegraphics[height=0.36\columnwidth]{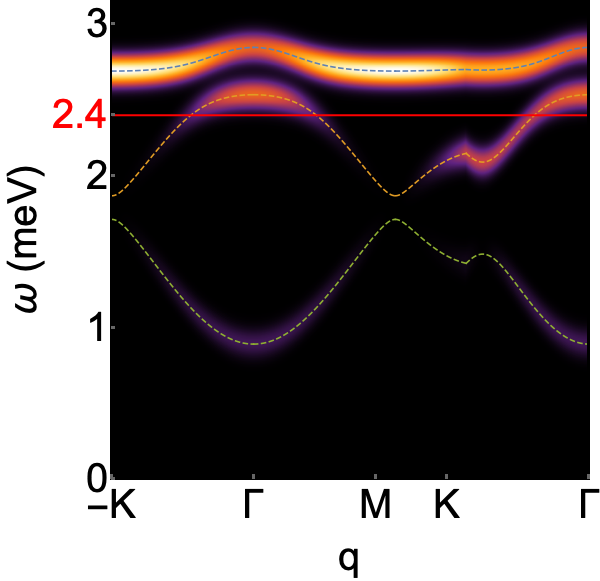}}
	\subfloat[Half moons in $S({\bm q},\omega  = 2.4\ \text{meV})$ \label{fig:4c}]{\includegraphics[height=0.36\columnwidth]{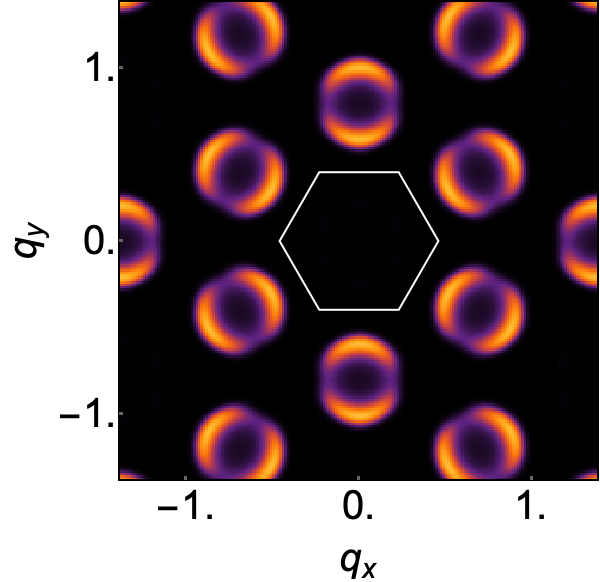}}
	\raisebox{5ex}{\includegraphics[width=0.07\columnwidth]{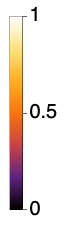}}
	\subfloat[Angle--integrated structure factor $S(|{\bm q}|,\omega)$]{\includegraphics[height=0.36\columnwidth]{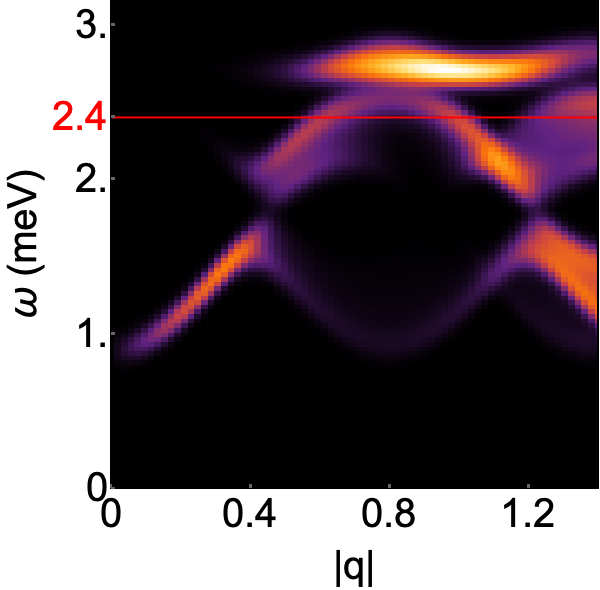}}	
	\subfloat[Half moons in $S(|\bm{q}|,\omega  = 2.4\ \text{meV})$ \label{fig:angle.integrated.spectrum}]{\includegraphics[height=0.36\columnwidth]{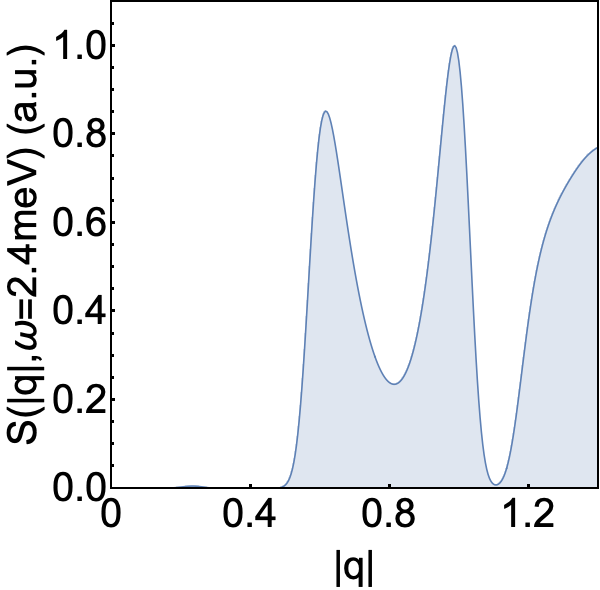}}	
	\caption{
		Half--moon features signalling topological bands in the Kagome 
		lattice ferromagnet Cu[1,3-bdc].
		(a) Magnon band structure for parameters taken from fits to 
		experiment \cite{Chisnell2015}, showing (approximately) flat
		band at high energies, and gaps to dispersing bands at lower energies.
		Because of the FM sign of $J$, magnon bands are inverted relative 
		to predictions shown in Fig~\ref{fig:HAF.LSW}.
		(b) Corresponding dynamical structure factor $S({\bm q}, \omega)$, 
		plotted on an irreducible wedge of the Brillouin Zone.
		(c) Half--moon features associated with topological band at intermediate energy, 
		as revealed by 
		$S({\bm q}, \omega = 2.4\ \text{meV})$.
		(d) Angle--integrated structure factor $S(|\bm{q}|,\omega)$, for 
		comparison with experiment \cite{Chisnell2015}. 
		(e) $S(|\bm{q}|,\omega  = 2.4\ \text{meV})$, showing double 
		peak as a consequence of the half--moon features.
		Results were calculated within linear spin wave theory for Eq.~(\ref{eq:H}), 
		with \mbox{$J = 0.6\ \text{meV}$}, \mbox{$D_z = 0.09\ \text{meV}$}, 
		\mbox{$g_z = 2.2$} and \mbox{$h = 7\ \text{T}$}. 
		Structure factors have been convoluted with a Gaussian of FWHM $0.15\ \text{meV}$
		to mimic experimental resolution.
	}
	\label{fig:kagome.FM}
\end{figure*}


\subsection{Inference of Berry Curvature from spectral features}
We now address the question of what we could have inferred about the 
topology of magnon bands 
from universal features in $S({\bm q},\omega)$ alone.
In a model with only two Magnon bands,  
the observation of half--moons approaching an avoided band--touching 
would be enough to completely characterise band topology, 
up to the sign of Chern numbers $C=\pm1$.
However the Kagome--lattice model we consider supports three 
band of magnons, and we must also take into account contributions 
to Berry phase coming from the third band; in particular the linear 
band--crossings at $K$ and $K'$.


The universal features associated with linear band--crossings 
are already well--characterised, comprising 
a single arc in $S({\bm{q}},\omega)$, which swaps orientation above
and below the (avoided) band--touching point~\cite{Shivam2017-arXiv}.
This phenomenology can be found in LSW calculations for 
Eq.~(\ref{eq:H}) \cite{supplemental-material}, and it is possible to infer 
Berry curvature $\pm \pi$ at $K$ and $K'$ from spectral features near these points.


Combining these results, we can safely infer the existence of 
topological magnon bands in the Kagome lattice model, Eq.~(\ref{eq:H}), 
from spectral features alone \footnote{An alternative line of reasoning, 
leading to  the same conclusion, is given in Ref.~\onlinecite{Thomasen2021-Thesis}.}. 
None the less, the sign of the contribution to Berry phase coming from each 
band--touching/bond--crossing point depends on details of interactions, and 
cannot be inferred solely from the existence of half--moons/arcs 
in $S({\bm{q}},\omega)$.
For this reason, additional information is required to assign a unique set 
of Chern numbers to these bands.
This could sought from thermal Hall measurements, which are sensitive 
to the sign of Berry curvature \cite{Katsura2010,Matsumoto2011} or, where models are 
sufficiently well--constrained, from fits to non--universal features in scattering.
 
\section{Application to experiment}
With these results in mind, it is interesting to revisit existing scattering data for frustrated magnets, 
and ask what we learn about possible topological band structures in these materials.
One example is \mbox{Cu(1,3-bdc)} which realises a structurally--perfect Kagome lattice \cite{Nytko2008}.
Neutron scattering experiments on \mbox{Cu(1,3-bdc)} \cite{Chisnell2015} 
reveal a ferromagnetic (FM) ground state,  and dispersing magnon bands. 
These are well--described by model with first--neighbour FM exchange 
and DM interactions, known to support a topological band structure.
In Fig.~\ref{fig:kagome.FM} we present LSW predictions for \mbox{Cu(1,3-bdc)}, 
for parameters taken from experiment.
Half--moon features are resolved near to an avoided band--touching 
at \mbox{$\omega \sim 2.5\ \text{meV}$}, shown in Figs.~\ref{fig:kagome.FM}(b,c).

Published spectra for \mbox{Cu(1,3-bdc)} were measured on powder samples \cite{Chisnell2015}.
In this case, half moons manifest as a double peak in predictions for the 
angle--integrated structure factor $S(|{\bm{q}}|,\omega)$ at fixed energy 
[Fig.~\ref{fig:angle.integrated.spectrum}].
This double peak originates in the way in which annuli of different $|{\bf q}|$ 
intersect half--moon features in reciprocal space [Fig.~\ref{fig:4c}], and for 
this specific crystal structure, is a robust consequence of the local constraint.
More generally, however, angle-resolved measurements of $S({\bm{q}},\omega)$ are 
preferable, since they eliminate any potential uncertainty about the 
origin of a double peak. 

 
Unambiguous observations of half--moon and pinch--point features 
already exist for a number of pyrochlore magnets.    
Perhaps the clearest example is found in Nd$_2$Zr$_2$O$_7$ \cite{Petit2016,Benton2016,Yan2018}, 
where half moons are associated with a quadratic band touching.
A very similar phenomonology is observed in Nd$_2$Hf$_2$O$_7$ \cite{Samartzis2022}.
Band topology was not previously discussed in either case, but from the available spectral evidence, 
we can classify both systems as at least topologically critical.
An example of a system known to posses a gapped, topological band structure is 
Lu$_2$V$_2$O$_7$ \cite{Zhang2013,Mena2014}.
In this case, half moon features are also observed above an avoided quadratic 
band touching \cite{Mena2016-Thesis}.
We anticipate that a similar phenomenology should be observed in the bilayer breathing 
Kagome magnet \CCO, for which single crystals are available \cite{Balz2016}.
In this case, published models of the high--field saturated state support flat magnon bands 
with a quadratic band touching \cite{Balz2017,Pohle2021}, and a topological band 
structure is also to be expected. 
 
Another interesting avenue to explore would be correlations 
within topological bands of electrons.
The phenomenology of half moons and Chern numbers developed in this 
work applies equally to electrons, and ``Kagome metals'' such as Fe$_3$Sn$_2$ [\onlinecite{Ye2018}] and CoSn [\onlinecite{Kang2020}] are already under study;  similar topological bands are argued to be possible in conducting MOF's \cite{Yamada2016}.
In this context, it would be interesting to revisit, for example, ARPES experiments 
on CoSn [\onlinecite{Kang2020}], where a flat band is observed 
a little below the Fermi energy, and reinterpret data in the terms of 
[Eq.~(\ref{eq:S.q.omega})]~\footnote{\textcolor{red}{Since completing this work, we have become 
aware of a preprint which explores pinch points in photoemission spectra of 
kagome and pyrochlore metals~\cite{Udagawa-arXiv}.}}.


\section{Conclusions} 
Topological bands of excitations are a ubiquitous feature of systems with both 
itinerant and localised electrons, originating in the Berry curvature of the underlying 
band states.
In this work, we have shown that the existence of pinch--points and half--moons 
in dynamical structure factors approaching an (avoided) quadratic band--touching 
point guarantees a net source of Berry curvature $\pm 2\pi$. 
The link between Berry curvature and spectral features is provided by the effective two--level 
model, Eq.~(\ref{eq:Dirac}), which describes the coupling between excitations with 
incompressible and irrotational character, and is characterised by a vector 
${\bm d}$ [Eq.~(\ref{eq:d})], that winds twice around the band--touching point.
These results imply a robust connection between singular features in scattering 
and Berry curvature  [Fig.~\ref{fig:summary}] which can, in some cases, be used to 
unambiguously determine the topological nature of bands from spectral features alone.
This picture is applicable to magnon bands in a wide range of frustrated magnets, 
and is also relevant to topological bands of electrons on frustrated lattices. 


\indent
{\it Acknowledgements}.  
This work was supported by the Theory of Quantum Matter Unit, Okinawa Institute 
of Science and Technology Graduate University (OIST).
The authors would like to thank Bella Lake and Matthias Gohlke for helpful discussions.
H.Y. is supported by the Theory of Quantum Matter Unit at Okinawa Institute of Science 
and Technology, the Japan Society for the Promotion of Science (JSPS) Research Fellowships 
for Young Scientists, and the
National
Science Foundation Division of Materials Research under
the Award DMR-191751  at 
different stages of this project.
J. R. was supported by the NSF through grant DMR-2142554.
%

\bibliography{paper.bib}

\clearpage
\setcounter{equation}{0}
\setcounter{figure}{0}
\setcounter{table}{0}
\makeatletter
\renewcommand{\theequation}{S\arabic{equation}}
\renewcommand{\thefigure}{S\arabic{figure}}
\renewcommand{\bibnumfmt}[1]{[#1]}
\renewcommand{\citenumfont}[1]{#1}

\onecolumngrid

\begin{center}
	\large{\textbf{Supplemental Material}}
\end{center}

\section{Microscopic model }

\subsection{Complete model for D$_{6h}$ symmetry}

The microscopic model considered in this work is a spin--1/2 magnet with anisotropic exchange 
interactions on the first--neighbor bonds of a kagome lattice.
This lattice has point--group symmetry D$_{6h}$ \cite{Essafi2017,Thomasen2021-Thesis}, 
with the existence of a mirror plane 
restricting anisotropic exchange interactions to transverse components 
of spin components.
Given these contstraints, the most general model allowed by the symmetry of the lattice is given by
\begin{eqnarray}
	\mathcal{H} &=&
	J_\perp \sum_{\langle ij \rangle} \left(S^x_i S^x_j +S^y_i S^y_j \right) +J_z \sum_{\langle ij \rangle} S^z_i S^z_j
	+ D_z \sum_{\langle ij \rangle} (\bm{S}_i\times\bm{S}_j)_z  
	+ K_\perp  \sum_{\langle ij \rangle} \mathbf{n}_{ij} \cdot \mathbf{Q}^\perp_{ij}
	- g_z h^z\sum_i S^z_i \; ,
	\label{eq:H_ani}
\end{eqnarray}
where the sum $\langle ij \rangle$ runs over first--neighbor bonds, $h^z$ is a magnetic field applied perpendicular 
to the mirror plane, $\mathbf{n}_{ij}$ is a unit vector in the direction of the bond pointing from site $i$ to site $j$, and 
\begin{eqnarray}
	\mathbf{Q}^\perp_{ij} = 
	\left(\begin{array}{c}
		S^x_i S^x_j - S^y_i S^y_j\\ 
		S^x_i S^y_j + S^y_i S^x_j \end{array}
	\right)  \; .
\end{eqnarray} 


The conventions for labeling lattice sites with the unit cell, and for counting the sense of 
bonds for DM interactions, $D_z$, are defined in Fig.~2 of main text.
Except where comparing with experiment [Fig.~4 of main text], 
we set the lattice parameter $a=1$, and consider sites located at positions 
\begin{equation}
	\mathbf{r}_{\sf A} = \frac{1}{4}(-1,\sqrt{3}),\quad
	\mathbf{r}_{\sf B} = \frac{1}{4}(1,\sqrt{3}),\quad
	\mathbf{r}_{\sf C} = \frac{1}{2}(0,-1) \; ,
	\label{eq:ABC}
\end{equation}
relative to the center of the primitive unit cell.

\subsection{Alternative parameterization}

An alternative parameterisation of the model Eq.~(\ref{eq:H_ani}) is used  
in \cite{Essafi2017,Thomasen2021-Thesis}.
In this approach, the interactions on first--neighbor bonds are most conveniently expressed as 
\begin{eqnarray}
	\mathcal{H} = \sum_{\langle ij \rangle}  \mathbf{S}_i J_{ij}  \mathbf{S}_j  - h^z\sum_{i} S^z_i \; ,
	\label{eq:4paramham}
\end{eqnarray}
where $J_{ij}$ is a tensor which depends on the sub--lattice of sites $i$ and $j$.     
Labelling sites according to the conventions of Eq.~(\ref{eq:ABC}) [cf. Fig.~1 of main text], this tensor is given by
\begin{subequations}
	\begin{eqnarray}
		J_{\sf AB} &=& 
		\begin{bmatrix}
			J_x&D_z&0\\
			-D_z&J_y&0\\
			0&0&J_z
		\end{bmatrix}\\
		J_{\sf BC} &=& 
		\begin{bmatrix}
			\frac{1}{4}(J_x+3J_y)& \frac{\sqrt{3}}{4}(J_x-J_y) + D_z&0\\
			\frac{\sqrt{3}}{4}(J_x-J_y)-D_z& \frac{1}{4}(3J_x+J_y)&0\\
			0&0&J_z
		\end{bmatrix}\\
		J_{\sf CA} &=& 
		\begin{bmatrix}
			\frac{1}{4}(J_x+3J_y)& -\frac{\sqrt{3}}{4}(J_x-J_y) + D_z&0\\
			-\frac{\sqrt{3}}{4}(J_x-J_y)-D_z& \frac{1}{4}(3J_x+J_y)&0\\
			0&0&J_z
		\end{bmatrix} \; , 
	\end{eqnarray}
\end{subequations}
Comparing the parametrization of Eq.~(\ref{eq:H_ani}) with that of Eq.~(\ref{eq:4paramham}), we see that 
\begin{eqnarray}
	J_\perp =  \frac{J_x + J_y}{2}         \quad , \quad K_\perp = \frac{J_x - J_y}{2} \; .
\end{eqnarray}

\section{Magnon bands within linear spin wave theory}
\label{sec:LSW}

\subsection{Spinwave Hamiltonian}

The results shown in Fig.~3 and Fig.~4 of the main text were calculated within linear spinwave (LSW) 
theory for Eq.~(\ref{eq:H_ani}), starting from a ground state fully polarised by either magnetic field [Fig 3], 
or ferromagnetic exchange interactions [Fig 4].
The fully--polarised state supports three bands of magnon excitations.
Within LSW theory, these bands are characterized by the Hamiltonian 
\begin{eqnarray}
	\mathcal{H}_{\sf LSW}
	= \sum_{\bf k} \left(
	\begin{array}{c}
		{\bm a}^\dagger_{\bf k}  \\
		{\bm a}^{\phantom{\dagger}}_{-{\bf k}}
	\end{array}
	\right)^T\!\!
	\left(\begin{array}{cc}
		{\bm M}_{\bf k} & {\bm N}_{\bf k} \\
		{\bm N}^*_{-{\bf k}} &  {\bm M}^*_{-{\bf k}} 
	\end{array}
	\right)
	\left(
	\begin{array}{c}
		{\bm a}^{\phantom{\dagger}}_{\bf k}  \\
		{\bm a}^{\dagger}_{-{\bf k}}
	\end{array}
	\right)\;,
	\label{eq:H.LSW}
\end{eqnarray}
where 
\begin{equation}
	{\bm a}^\dagger_{\bf k}=(a^\dagger_{{\sf A},\bf k},a^\dagger_{{\sf B},\bf k},a^\dagger_{{\sf C},\bf k}) \,
\end{equation}
describes magnons created on the sublattice $\{ {\sf A}, {\sf B}, {\sf C} \}$, with dispersion determined by block--diagonal,  
magnon--hopping terms
\begin{eqnarray}
	{\bm M}_{\bf k} =\left(\begin{array}{ccc}
		g_z h^z -2J_z   & (J_\perp +i D_z)\cos\frac{{\bm \delta}_{\sf AB}\cdot{\bf k}}{2} & (J_\perp -I D_z)\cos\frac{{\bm \delta}_{\sf CA}\cdot{\bf k}}{2} \\
		(J_\perp -i D_z)\cos\frac{{\bm \delta}_{\sf AB}\cdot{\bf k}}{2}   &  g_z h^z -2J_z & (J_\perp +i D_z)\cos\frac{{\bm \delta}_{\sf BC}\cdot{\bf k}}{2}\\
		(J_\perp +i D_z)\cos\frac{{\bm \delta}_{\sf CA}\cdot{\bf k}}{2} & (J_\perp -i D_z)\cos\frac{{\bm \delta}_{\sf BC}\cdot{\bf k}}{2}&  g_z h^z -2J_z 
	\end{array}
	\right)\;,
	\label{eq:LSW.matrix.M}
\end{eqnarray}
which conserve magnon number, and off--diagonal, magnon--pairing terms 
\begin{eqnarray}
	{\bm N}_{\bf k} =\left(\begin{array}{ccc}
		0   & K_\perp \cos\frac{{\bm \delta}_{\sf AB}\cdot{\bf k}}{2} & K_\perp e^{i\frac{2\pi}{3}} \cos\frac{{\bm \delta}_{\sf CA}\cdot{\bf k}}{2} \\
		K_\perp \cos\frac{{\bm \delta}_{\sf AB}\cdot{\bf k}}{2}   &  0 & K_\perp e^{-i \frac{2\pi}{3}} \cos\frac{{\bm \delta}_{\sf BC}\cdot{\bf k}}{2}\\
		K_\perp e^{i \frac{2\pi}{3}} \cos\frac{{\bm \delta}_{\sf CA}\cdot{\bf k}}{2} & K_\perp  e^{-i \frac{2\pi}{3}}\cos\frac{{\bm \delta}_{\sf BC}\cdot{\bf k}}{2}&  0 
	\end{array}
	\right) \; .
	\label{eq:LSW.matrix.N}
\end{eqnarray}
which depend solely on the symmetric part of the exchange anisotropy, $K_\perp$.
Since the interactions in Eq.~(\ref{eq:H_ani}) are restricted to first--neighbor bonds, lattice Fourier transforms depend only 
on the vectors linking neighbouring sites
\begin{equation}
	{\bm \delta}_{\rm ij} = a \mathbf{n}_{ij} = \mathbf{r}_i - \mathbf{r}_j \; ,
\end{equation}
with $\{\mathbf{r}_i, \mathbf{r}_j\}$ defined through  Eq.~(\ref{eq:ABC}).


\begin{figure}[ht!]
	\centering
	\subfloat[\label{Fig.Kagome.SW.dispersion.fin.K}  Band dispersion, \hspace{3.5cm} $K_\perp = 0.63, D_z = 0$
	]{\includegraphics[height=6cm]{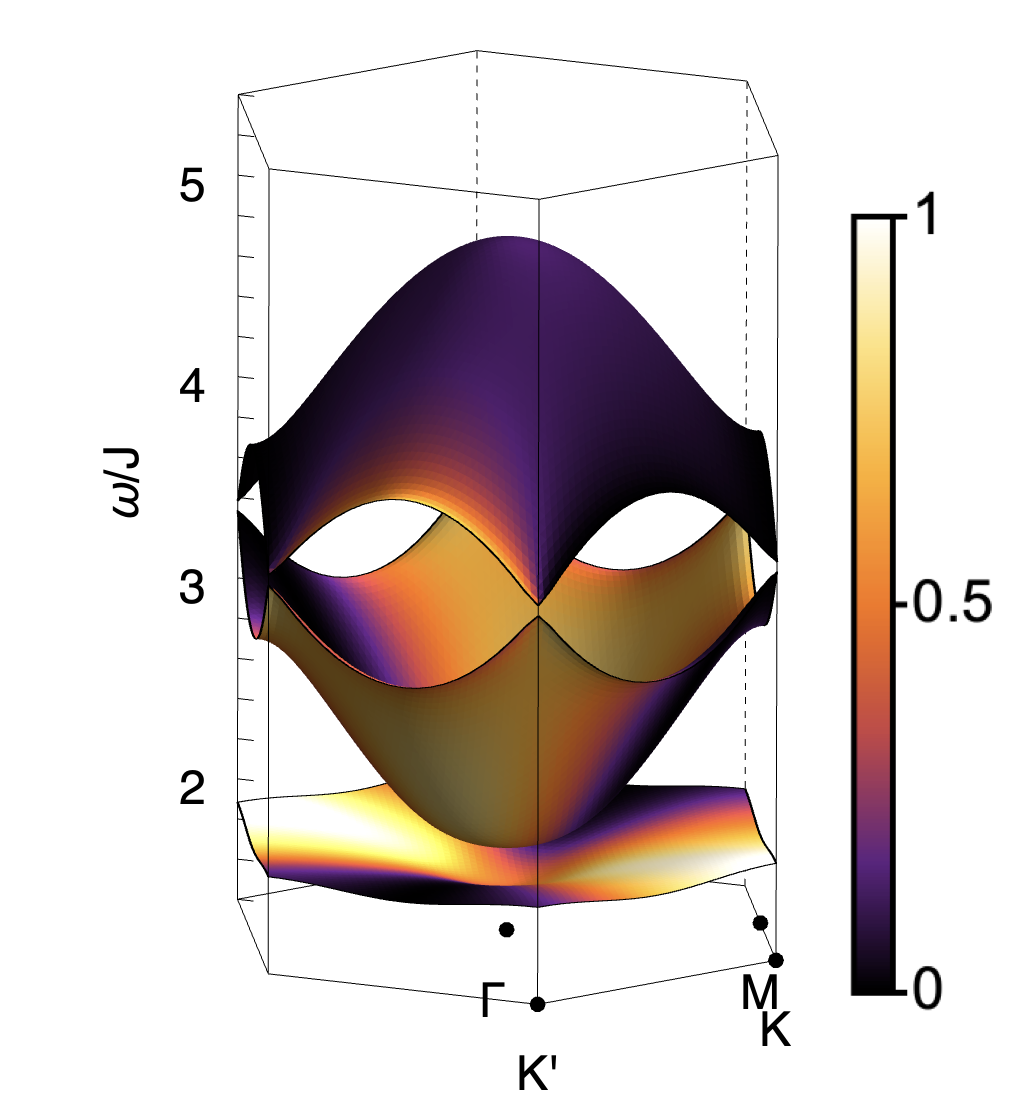}}
	\subfloat[\label{Fig:Berry.curvature.fin.K}  Berry curvature, \hspace{3.5cm} $K_\perp = 0.63, D_z = 0 $ ]{\includegraphics[height=6cm]{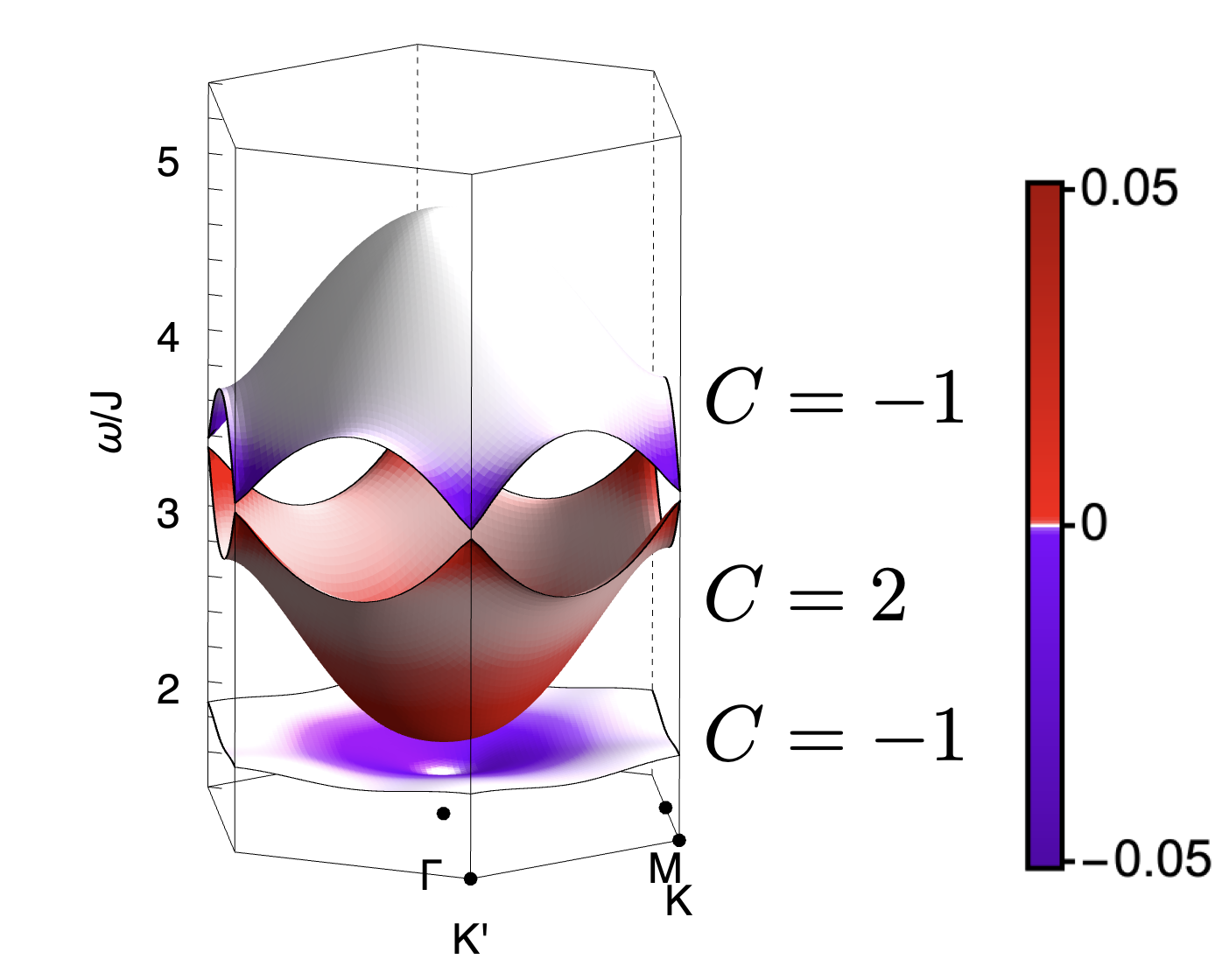}  }\\
	\subfloat[\label{Fig.Kagome.K.SW.1}$S({\bf k}, \omega = 2.02)$, \hspace{3.5cm}  $K_\perp = 0.63, D_z = 0$
	]{\includegraphics[height=4.5cm]{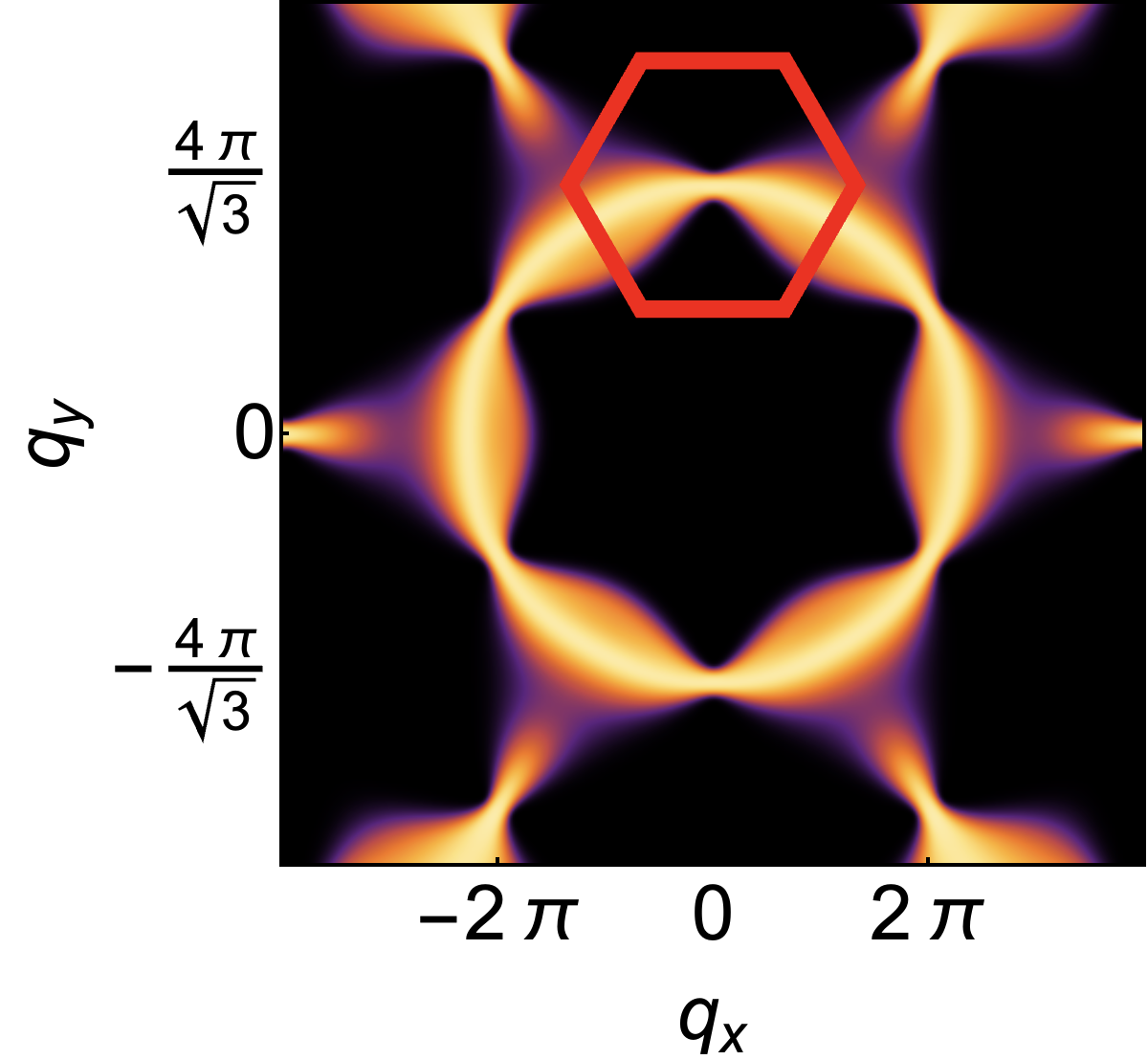}}\ \ 
	\subfloat[\label{Fig.Kagome.K.SW.2}$S({\bf k}, \omega =  2.04)$, \hspace{3.5cm}  $K_\perp = 0.63, D_z = 0$
	]{\includegraphics[height=4.5cm]{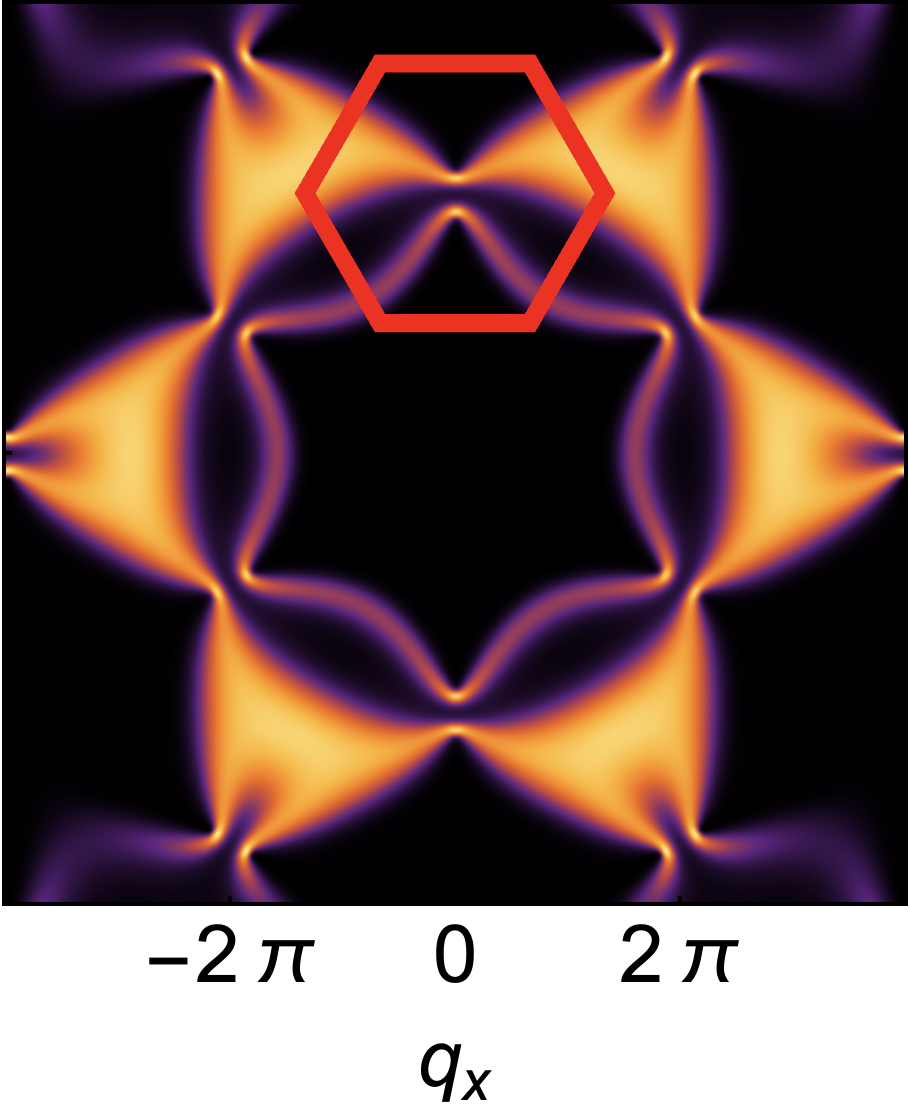}}\ \ 
	\subfloat[\label{Fig.Kagome.K.SW.3}$S({\bf k}, \omega = 2.19)$,\hspace{3.5cm}  $K_\perp = 0.63, D_z = 0$
	]{\includegraphics[height=4.5cm]{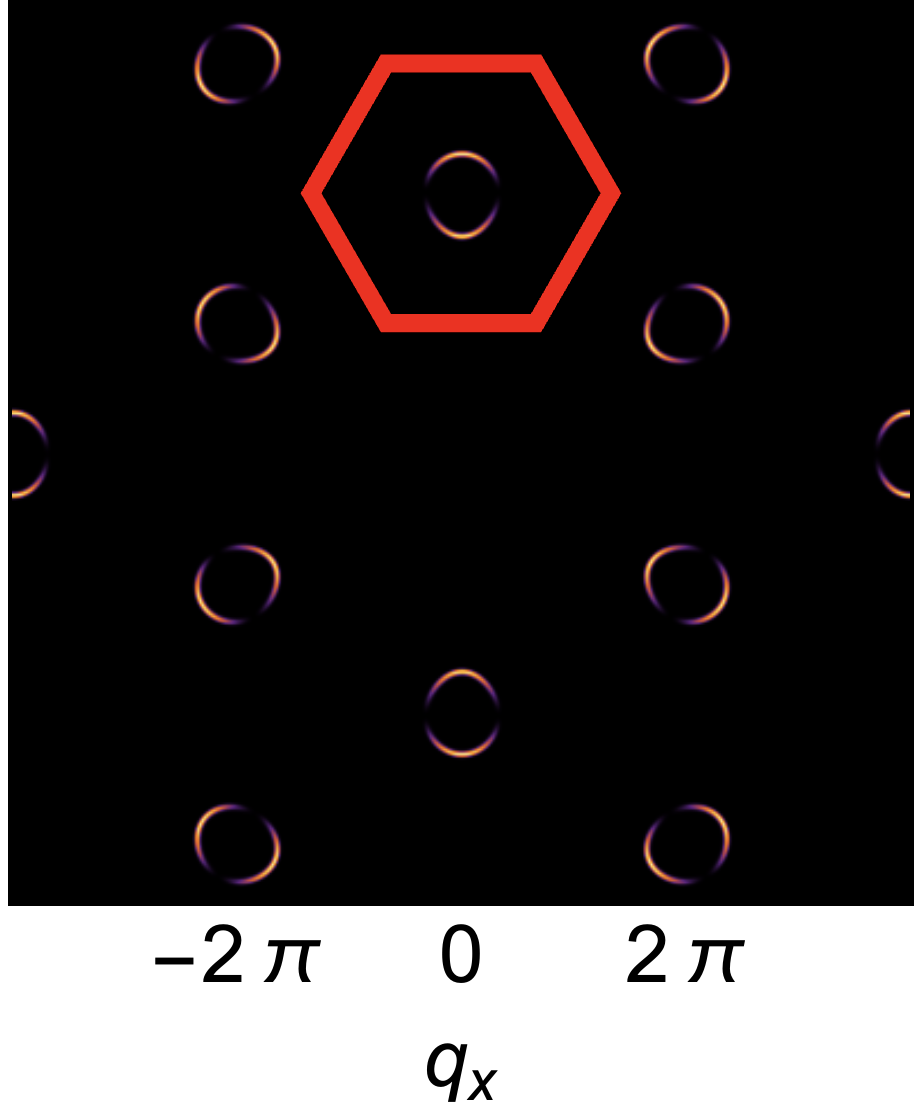}}
	\caption{
		Relationship between spectral features, Berry curvature and band topology in a model with bond--symmetric exchange anistoropy, $K_\perp$. 
		(a)~Spin--wave dispersion, illustrating how the mixing of states between bands opens 
		gaps at band--touching points.
		The colorscale shows how each band contributes to $S({\bf k}, \omega)$,
		(b)~Berry curvature associated with spin--wave bands.
		Integrated accross the Brillouin zone (BZ), this leads to bands with the Chern numbers $C = -1,~2,~-1$.
		(c)~Dynamical structure factor $S({\bf k}, \omega)$ at $\omega = 2.02$, 
		showing how the mixing of states between bands eliminates the pinch--point singularity for 
		${\bf k} \to \Gamma$.
		Red hexagon denotes the BZ considered in (a)--(b).
		(d)~Equivalent results for $\omega = 2.04$, showing the change of the structure factor pattern.
		(e)~Equivalent results for $\omega = 2.19$, showing   survival of half moons away from 
		zone center in the middle band. 
		All results were obtained within linear spin wave (LSW) theory for 
		Eq.~\eqref{eq:H_ani}, for parameters $J_z = J_\perp  = 1$, $g_zh^z = 5$, $K_\perp = 0.63,\ D_z =0$.
		Results in (c)--(e) have been convoluted with a Gaussian of FHWM = $0.03\ J_z$ 
		to mimic experimental resolution.
	} 
	\label{fig:HAF.LSW.K}
\end{figure}


The spin wave spectrum is determined by solving the Bogoliubov-de Gennes equation
\begin{eqnarray}
	\left(\begin{array}{cc}
		{\bm M}_{\bf k} & {\bm N}_{\bf k} \\
		{\bm N}^*_{-{\bf k}} &  {\bm M}^*_{-{\bf k}} 
	\end{array}\right) 
	\left| n_\lambda ({\bf k}) \right>
	= \omega_\lambda (\mathbf{k}) \Sigma^z 
	\left| n_\lambda ({\bf k}) \right> \;,
	\label{eq:BdG}
\end{eqnarray}
where
\begin{eqnarray}
	\Sigma_z  = \left(\begin{array}{cc}
		{\bm 1} & {\bm 0} \\
		{\bm 0}&  -{\bm 1}
	\end{array}
	\right) \; .
\end{eqnarray}
%
	%
	%
	In practice this solution is implemented by multiplying the spin--wave Hamiltonian $\mathcal{H}_{\sf LSW} ({\bf k})$ 
	by the pseudo--unitary matrix $\Sigma^z$, and numerically diagonalising the resulting matrix \cite{Balian1969}
	\begin{eqnarray}
		\overline{\bm H}_{\sf LSW} 
		=
		\left(\begin{array}{cc}
			{\bm M}_{\bf k} & {\bm N}_{\bf k} \\
			-{\bm N}^*_{-{\bf k}} &  -{\bm M}^*_{-{\bf k}} 
		\end{array}\right) \; .
		\label{eq:H.LSW.effective}
	\end{eqnarray}
	Once the eigenvectors $ \left| n_\lambda ({\bf k}) \right>$ are known, the Berry curvature of bands can be 
	calculated as 
	\begin{eqnarray}
		F^{xy}_\lambda ({\bf k}) 
		= i\left< \partial_{k_x} n_\lambda ({\bf k})| \Sigma^z \partial_{k_y} n_\lambda({\bf k})\right>
		- i \left< \partial_{k_y} n_\lambda ({\bf k}) |\Sigma^z \partial_{k_x} n_\lambda({\bf k})\right> \; .
		\label{eq:Berry_general}
	\end{eqnarray}
	This can in turn be integrated over the Brillouin Zone to obtain integer Chern numbers 
	\begin{eqnarray}
		C_\lambda  = \frac{1}{2\pi i} \int_{\rm BZ} dk_x dk_y F^{xy}_\lambda ({\bf k}) \;.
	\end{eqnarray}
	These characterise band topology.   
	Further details of this type of calculation can be found in \cite{Thomasen2021}.
	
	
	It is the solution of Eq.~\ref{eq:BdG}, evaluated for $J_\perp = J_z = J$, $K_\perp = 0$, 
	which determines the dispersions, structure factors and Berry curvature plotted in 
	Fig.~3 and Fig.~4 of the main text.
	Results for for a finite value of $K_\perp$, showing equivalent pinch--point and half moon 
	features, are shown in Fig.~\ref{fig:HAF.LSW.K}.
	
	\subsection{Phase diagram: band topology as a function of $K_\perp$ and $D_z$ }
	\label{sec:phase.diagram}
	
	
	\begin{figure}[t]
		\centering
		\includegraphics[height=5cm]{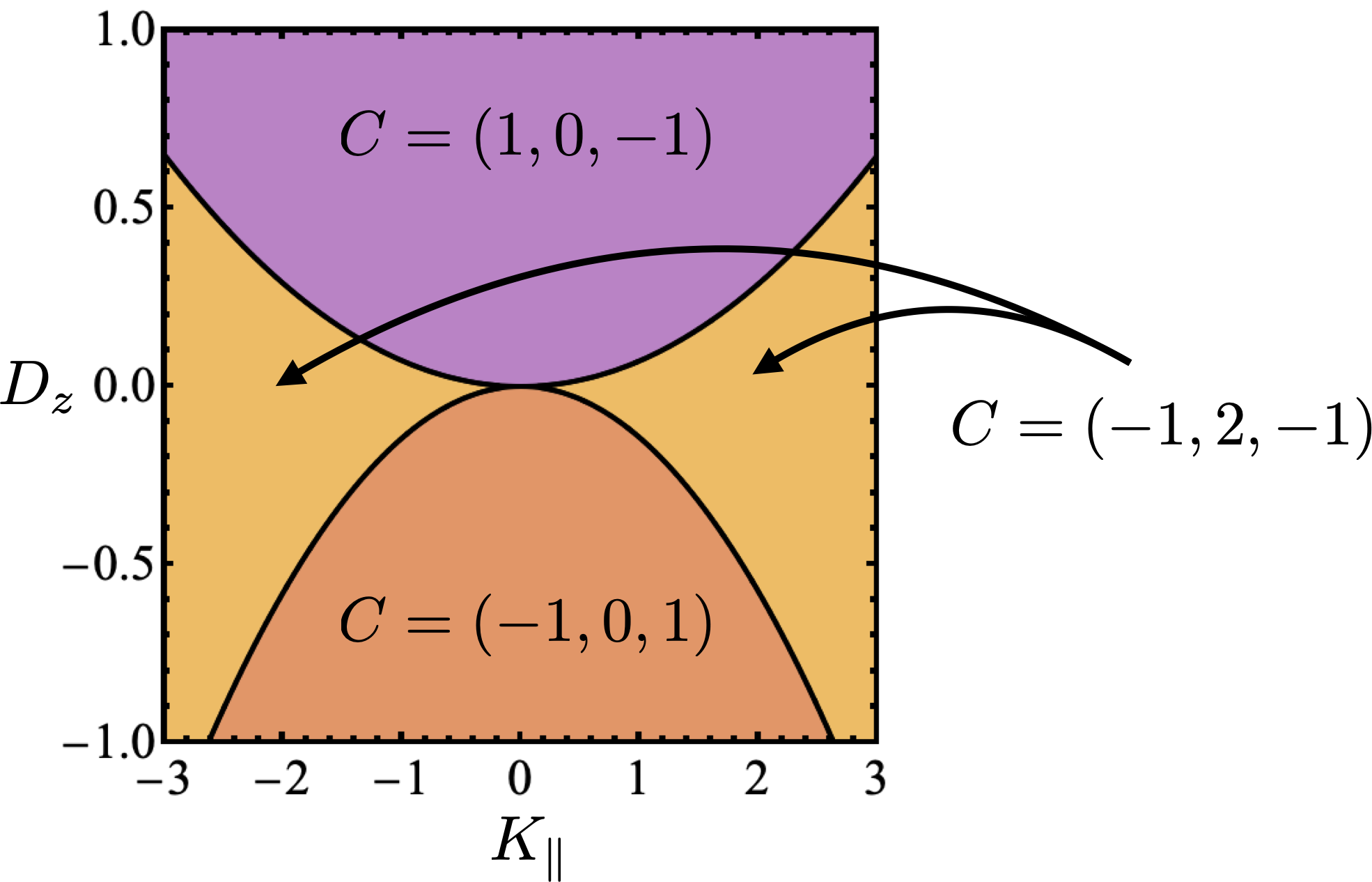}
		\caption{ 
			Phase diagram of the Kagome--lattice model model, Eq.~(\ref{eq:H_ani}), in high magnetic field, 
			as a fucntion of Dzyaloshinskii--Moriya (DM) interactions, $D_z$ and bond--symmetric interaction, 
			$K_\perp$, showing the Chern numbers of associated magnon bands.   
			Because $K_\perp$ enters band structures as $\sim K_\perp^2/h_z$ [Section~\ref{sec:K.perp}], 
			its contribution is weaker than that of DM interactions, and independent of the sign of $K_\perp$.
			Results are shown for $J_z = J_\perp  = 1$, and magnetic field $h^z = 5$, sufficient to fully 
			polarize spins.
			The numbers in each phase are the Chern numbers for the top, middle, and bottom 
			band respectively [cf. Fig.~(\ref{fig:HAF.LSW.K})] and Fig.~3 of main text].
			The role of band crossings in determining these Chern numbers is illustrated in Fig.~\ref{FIG_gap_open}.
		} 
		\label{fig:phase.diagram}
	\end{figure}
	
	
	\begin{figure}[ht!]
		\centering
		\subfloat[\label{FIG_dispersion_C_Dz_0}Band dispersion and  Chern number, \hspace{3.5cm} $K_\perp = 0, D_z = 0$ 
		]{\includegraphics[height=4cm]{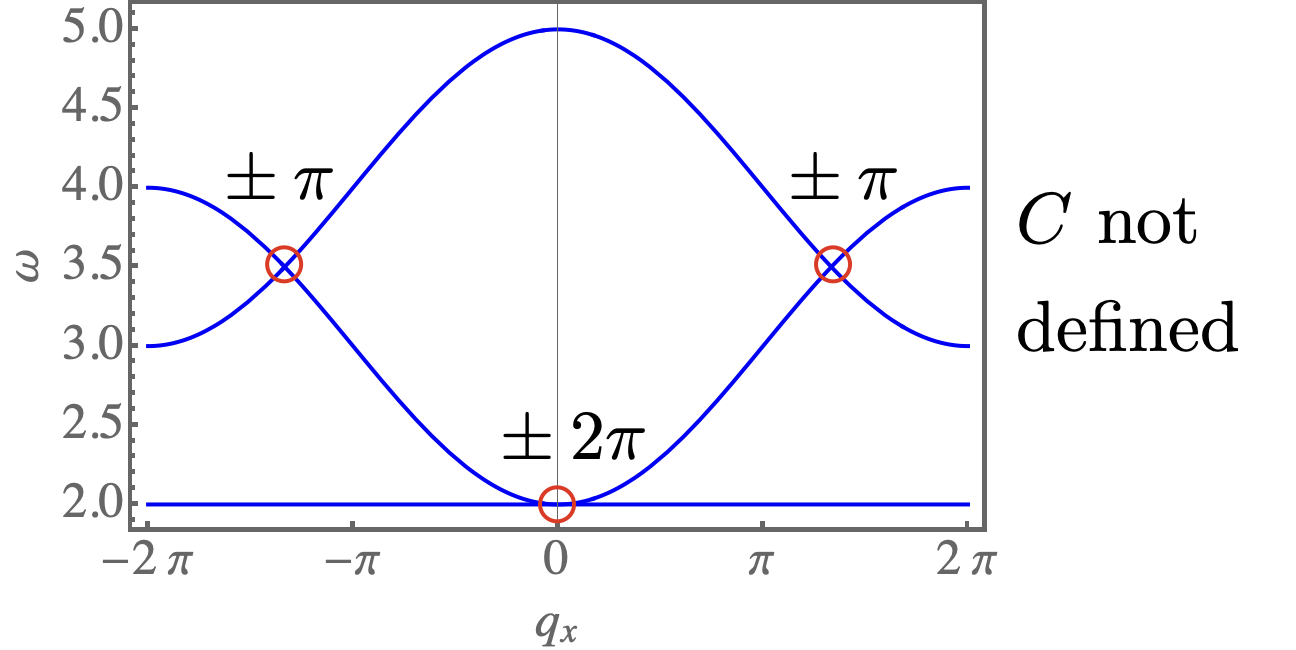}}
		\subfloat[\label{FIG_dispersion_C_Dz_0_1} Band dispersion and  Chern number, \hspace{3.5cm} $K_\perp = 0, D_z = 0.1$ ]{\includegraphics[height=4cm]{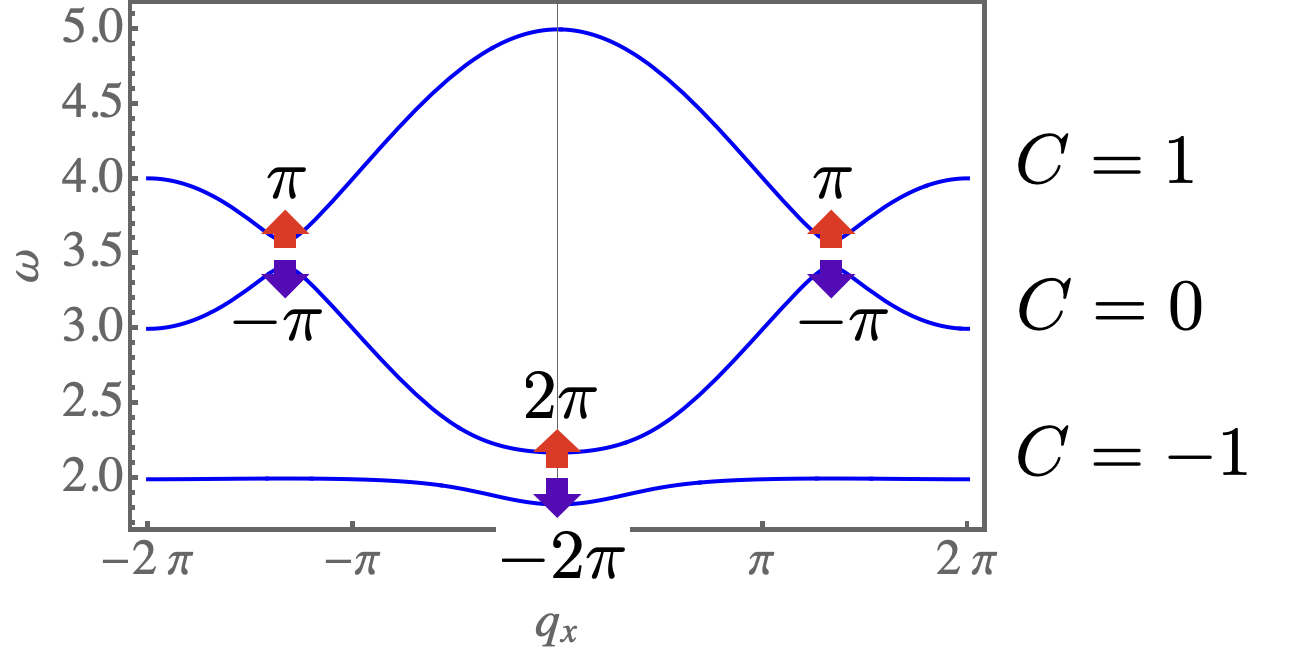}  } \\
		\subfloat[\label{FIG_dispersion_C_Dz_}Band dispersion and Chern number, \hspace{3.5cm} $K_\perp = 0, D_z = -0.1$ 
		]{\includegraphics[height=4cm]{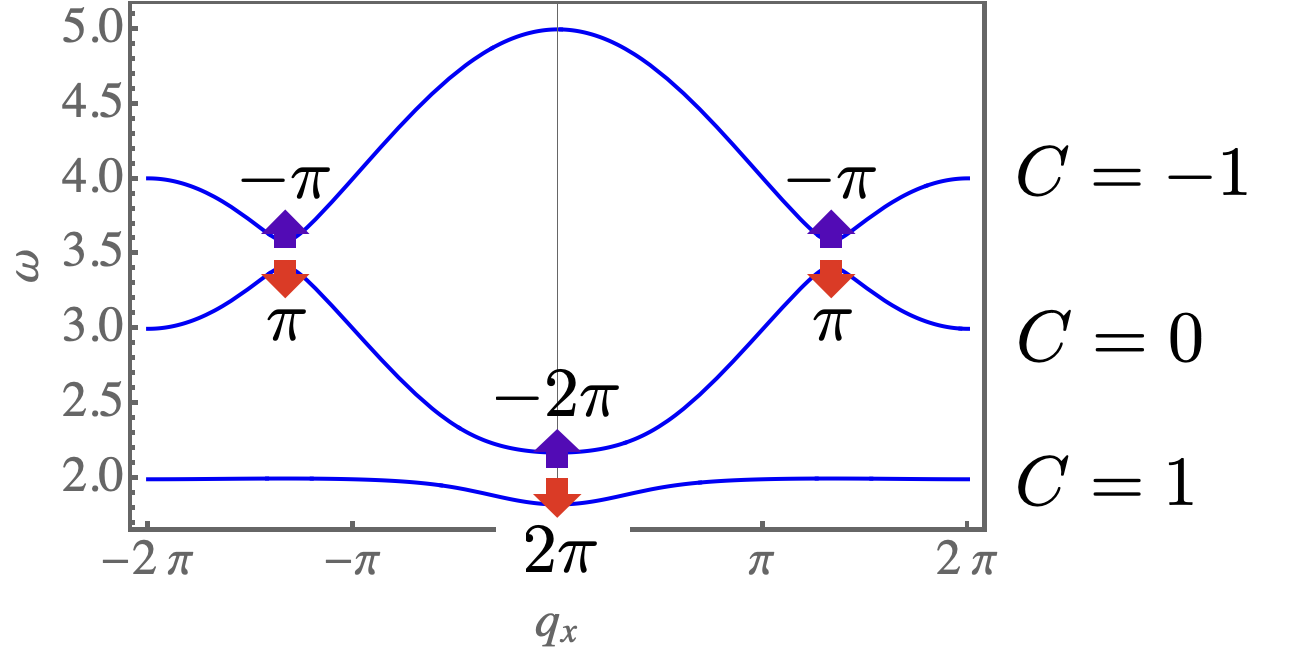}}
		\subfloat[\label{FIG_dispersion_C_K_1} Band dispersion and Chern number, \hspace{3.5cm} $K_\perp = 1, D_z = 0$ ]{\includegraphics[height=4cm]{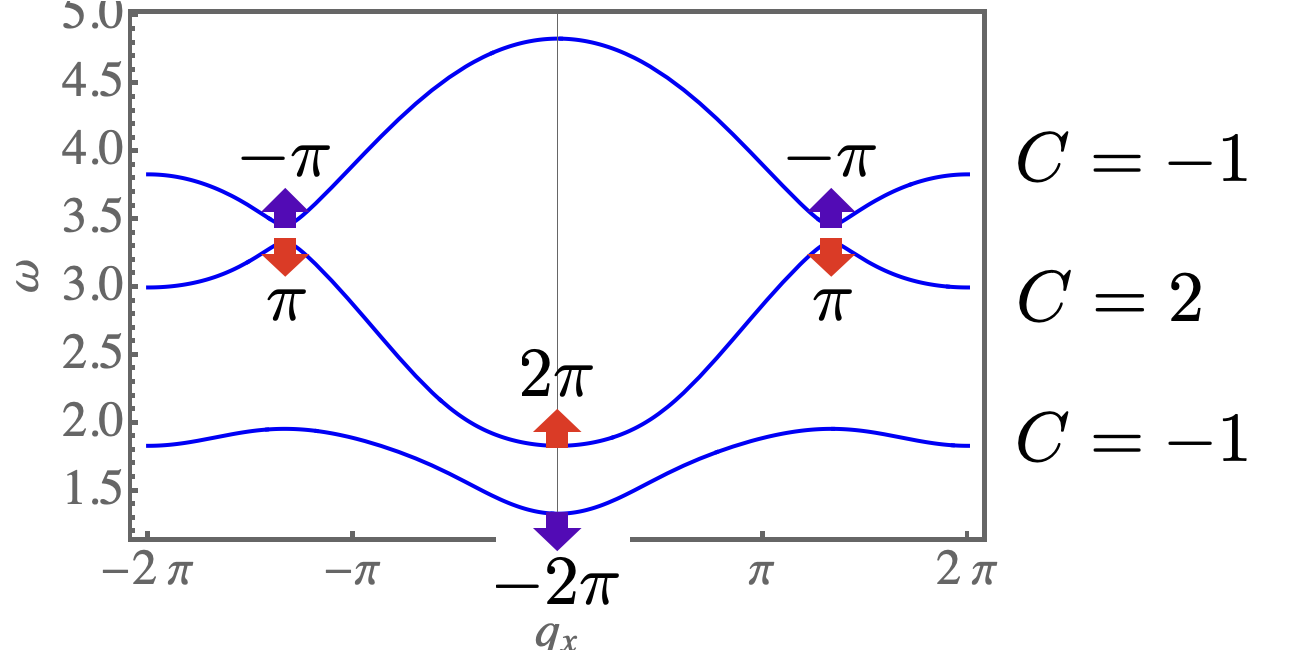}  } 
		
		\caption{
			Magnon band topologies, and their relationship with band--touching points.
			(a) Magnon band structure for vanishing bond--symmetric anisotropic exchange 
			$K_\perp=0$ and DM interaction $D_z=0$.
			A Berry phase of $\pm\pi$ is associated with each of the linear band crossings (Dirac points) 
			on the zone boundary.
			Berry phase of $\pm2\pi$ is associated with the quadratic band--touching in the zone center.
			(b) Magnon band structure for DM interaction $K_\perp=0$, $D_z=0.1$, showing how 
			contributions from band--touching points contribute to integer Chern numbers. 
			(c) Magnon band structure for DM interaction $K_\perp=0$, $D_z=-0.1$, showing how Chern numbers 
			change sign relative to $D_z > 0$.
			(d) Magnon band structure for $K_\perp = 1 $, $D_z=0$, showing the alternative set 
			of Chern numbers following from $K_\perp$.
			The domain applicable to each of these cases is shown in Fig.~\ref{fig:phase.diagram}.
		} 
		\label{FIG_gap_open}
	\end{figure}
	
	
	For sufficiently large magnetic field, $h_z$, the ground state of Eq.~(\ref{eq:H_ani}) is fully 
	polarised, and posses three bands of magnon excitations.
	In the presence of finite DM interactions, $D_z$, or bond--symmetric interactions, $K_\perp$, 
	these bands have non--trivial topology.   
	In what follows we explore the different band topologies which arise as a function of 
	$D_z$ and $K_\perp$.    
	For simplicity, we set $J_\perp = J_z = J$.
	Results are summarized in Fig.~\ref{fig:phase.diagram}.
	
	
	When $D_z = K_\perp = 0$, magnon bands exhibit a quadratic band--touching points at the zone center
	${\bf q} = \Gamma$, and linear band crossings (Dirac points) at the zone corners ${\bf q} = K, K'$.
	This band structure, illustrated in Fig.~\ref{FIG_dispersion_C_Dz_0} and Fig.~3 of the main text, is topologically critical: 
	a Berry phase of $\pm 2\pi$ is associated with the quadratic band touching, while a  
	Berry phase of $\pm \pi$ is associated with each of the Dirac points.
	However, these contributions are singular, with the contributions to Berry phase concentrated  
	at the band--touching points.
	And in the absence of a gap, Chern numbers for the individual bands remain ill--defined.
	
	
	The case with $D_z = K_\perp = 0$ represents a line of critical points of Eq.~(\ref{eq:H_ani}), 
	for which both $D_z$ and $K_\perp$ act as singular perturbations.
	An infinitesimal value of either $D_z$ or $K_\perp$  is sufficient to open gaps at all three  
	band--touching points, and endow the magnon bands with finite Chern numbers.
	In the case of finite $D_z > 0$ ($K_\perp=0$), the singular contributions to Berry phase coming 
	from the band--touching points are resolved so as to give Chern numbers 
	$C = 1,~0,~-1$, as illustrated in Fig.~\ref{FIG_dispersion_C_Dz_0_1}.
	This is the case discussed in the main text [cf. Fig.~3 of main text].
	Changing the sign of $D_z$ changes the sign of its contributions, without changing their relative phase, 
	leading to Chern numbers  $C = -1,~0,~1$.
	This case is illustrated in Fig.~\ref{FIG_dispersion_C_Dz_}.
	
	
	The bond--symmetric interaction $K_\perp$ also mixes states between different bands, opening 
	a gap at all three band--touching points, and leading to a non--trivial band topology.
	However in this case the relative phase of contributions from quadratic and linear band touchings 
	is reversed, leading to Chern numbers $C = -1,~2,~-1$, regardless of the sign of $K_\perp$.
	This case is  illustrated in Fig.~\ref{FIG_dispersion_C_K_1}, and discussed in more detail 
	in Section~\ref{sec:K.perp}.
	
	
	We can derive a general phase diagram for Eq.~(\ref{eq:H_ani}) as a function of $K_\perp$ 
	and $D_z$, by tracking where in parameter space the gaps between bands close. 
	Doing so defines lines in parameter space, where some (or all) of the contributions 
	to Berry curvature coming from band--touching points change sign, leading to a an 
	overall change in Chern numbers.
	For general $K_\perp$, $D_z$, the gap at the $\Gamma$ point is given by 
	\begin{equation}
		\label{EQN_Gamma_point_gap}
		\Delta_\Gamma = 2\left|\sqrt{3} D_z+\frac{3 K_\perp^2}{4(g_z h^z -2 J_z)} \right| \; .
	\end{equation}
	%
	%
	Meanwhile, the gap at the $K$ point is
	\begin{equation}
		\Delta_K 
		=  \left|
		\sqrt{3}D_z -\frac{3K_\perp^2}{8(g_z h^z - 2 J_z)}
		\right| \; .
	\end{equation}
	%
	%
	The two curves defined by 
	\begin{equation}
		\Delta_\Gamma  = 0;\quad \Delta_K = 0 \; , 
	\end{equation}
	form the phase boundaries of the phase diagram shown in Fig.~\ref{fig:phase.diagram}.
	
	\section{Role of bond--symmetric anisotropic exchange, $K_\perp$}
	\label{sec:K.perp}
	
	\subsection{General considerations}
	
	As discussed in Section~\ref{sec:phase.diagram}, the bond--symmetric exchange term, $K_\perp$, 
	endows the magnon bands with Chern numbers $(-1,2,-1)$, in the absence of DM interactions, $D_z=0$.
	The spectral features associated with (avoided) band touching points in this case are qualitatively 
	identical to those discussed for $D_z$ in the main text, and the inferences which can be made about 
	local contributions to Berry phase therefore remain unchanged.
	However the way in which $K_\perp$ enters into LSW theory, Eq.~(\ref{eq:BdG}), is quite different from $D_z$.  
	
	
	The DM interactions $D_z$ conserve the number of magnons, and therefore appear in the 
	magnon ``hopping'' term ${\bm M}_{\bf k}$ [Eq.~(\ref{eq:LSW.matrix.M})].   
	Meanwhile the bond--symmetric exchange term, $K_\perp$, introduces magnon--pairing 
	terms, characterised through the matrix ${\bm N}_{\bf k}$ [Eq.~(\ref{eq:LSW.matrix.N})].
	These pairing terms mix the physical (particle) and unphysical (hole) subspaces of 
	$\mathcal{H}_{\sf LSW}$ [Eq.~(\ref{eq:H.LSW})].
	%
	%
	Where pairing terms are small compared with the gap between particle and hole subspaces, i.e. 
	\begin{eqnarray}
		K_\perp \ll 2 \Delta \approx 2(h^z - 2 J_z - J_\perp) \; ,
		\label{eq:condition.on.Kperp}
	\end{eqnarray}
	it is possible to eliminate them from Eq.~(\ref{eq:BdG}) by a suitable perturbative 
	canonical transformation \cite{MacDonald1988}.
	This will have the effect of renormalising magnon hopping term 
	\begin{eqnarray}
		{\bm M}_{\bf k} \to \tilde{{\bm M}}_{\bf k} \; ,
	\end{eqnarray}
	and removing ${\bm N}_{\bf k}$ from the theory entirely.
	As a consequence, the form of the equations of motion (EoM) used to describe pinch--point and 
	half--moon features in the main text remain unchanged, along with the conclusions 
	drawn from them about the relationship between spectral features and Berry phase.
	
	In Section~\ref{sec:canonical} we provide details of the canoncial transformation needed 
	to project pairing terms onto the physical subspace of the LSW theory.
	
	In Section~\ref{sec:EoM} we derive a general EoM describing an (avoided) quadratic 
	band--touching point in the presence of both $D_z$ and $K_\perp$.
	
	\subsection{Perturbative canonical transformation}
	\label{sec:canonical}
	
	We take as a starting point the Hamiltonian $\overline{\bm H}_{\sf LSW}$ [Eq.~(\ref{eq:H.LSW.effective})], 
	which determines magnon band structure.   
	The eigenstates of $\overline{\bm H}_{\sf LSW}$ occur with both positive energy (physical subspace)
	and negative energy (unphysical subspace), and we seek a canonical transformation which will eliminate all  
	terms connecting these two subspaces, such that
	\begin{eqnarray}
		\tilde{\bm H}_{\sf LSW} 
		=
		\left(\begin{array}{cc}
			\tilde{\bm M}_{\bf k} & {\bm 0} \\
			{\bm 0}  &  -\tilde{\bm M}^*_{-{\bf k}} 
		\end{array}\right) \; .
	\end{eqnarray}
	Ideally, this canonical transformation would be carried out exactly as
	\begin{eqnarray}
		\tilde{\bm H}_{\sf LSW} = e^{-{\bm S}} \overline{\bm H}_{\sf LSW} e^{\bm S} \; ,
	\end{eqnarray} 
	where ${\bm S}$ is an appropriate block--off--diagonal matrix.
	In the absence of prior knowledge of this matrix, we proceed perturbatively, constructing 
	${\bm S}$ order--by--order in perturbation theory \cite{MacDonald1988}.
	
	
	Since the block-off-diagonal elements in ${\bm N}_{\bf k}$ are small compared to the diagonal terms 
	in ${\bm M}_{\bf k} $, Eq.~(\ref{eq:condition.on.Kperp}), the unitary transformation 
	$e^{-{\bm S}}$ must be close to the identity.
	We can therefore expand the transformation as a power series in ${\bm S}$
	\begin{eqnarray}
		e^{-{\bm S}}= {\bm 1} - {\bm S} +\frac 1 2 {\bm S}^2 -\frac{1}{3!} {\bm S}^3 +\hdots \; .
	\end{eqnarray} 
	It follows that the transformed spin-wave Hamiltonian is given by
	\begin{equation}
		\tilde{\bm H}_{\sf LSW} =e^{-{\bm S}} \overline{\bm H}_{\sf LSW}   e^{{\bm S}} 
		= \sum_n \frac{1}{n!} \left[\overline{\bm H}_{\sf LSW} , {\bm S} \right]^{(n)} \; , 
	\end{equation}
	with
	\begin{eqnarray}
		\left[ \overline{\bm H}_{\sf LSW} ,{\bm S} \right]^{(n+1)}= \left[[\overline{\bm H}_{\sf LSW} ,{\bm S}]^{(n)}, {\bm S} \right] 
		\quad \; , \quad 
		[\overline{\bm H}_{\sf LSW} ,{\bm S}]^{(0)}=\overline{\bm H}_{\sf LSW}  \; .
		\label{eq:canon}
	\end{eqnarray} 
	
	
	As the next step, we separate the block--off--diagonal and block--diagonal parts 
	of $\tilde{\mathcal{H}}_{\sf LSW}$. 
	We decouple the spin-wave Hamiltonian as  
	\begin{eqnarray}
		\overline{\bm H}_{\sf LSW} = {\bm H}_0+{\bm H}_1+{\bm H}_2 \; ,
	\end{eqnarray} 
	here ${\bm H}_0+{\bm H}_1$ is the block--diagonal part of  $H^{{\rm sw}}$, containing only 
	matrix elements of ${\bm M}$ [Eq.~(\ref{eq:H.LSW.effective})].
	Within this, ${\bm H}_0$ contains diagonal elements of $\overline{\bm H}_{\sf LSW}$, 
	while ${\bm H}_1$ collects all other block--diagonal terms. 
	Meanwhile, ${\bm H}_2$ is the block--off--diagonal part of $\overline{\bm H}_{\sf LSW}$, 
	composed exclusively of matrix elements of ${\bm N}_{\bf k}$ and ${\bm N}^*_{-{\bf k}}$.
	Under the canonical transformation, Eq.~(\ref{eq:canon}), these terms transform as  
	\begin{eqnarray}
		\tilde{\bm H}_{\text{off-diag}} &=&
		\sum_{n=0}^{\infty}\frac{1}{(2n +1)!}\left[{\bm H}_0+{\bm H}_1,{\bm S}\right]^{(2n+1)} 
		+\sum_{n=0}^{\infty}\frac{1}{(2n)!}\left[{\bm H}_2,{\bm S}\right]^{(2n)} \;,
		\label{eq:off_diag}
	\end{eqnarray}
	
	\begin{eqnarray}
		\tilde{\bm H}_{\text{diag}} &=& \sum_{n=0}^{\infty}\frac{1}{(2n)!}
		\left[{\bm H}_0 + {\bm H}_1,{\bm S}\right]^{(2n)}
		+\sum_{n=0}^{\infty}\frac{1}{(2n+1)!}\left[{\bm H}_2,{\bm S}\right]^{(2n+1)} \; .
		\label{eq:diag}
	\end{eqnarray} 
	We are now in a position to solve for ${\bm S}$ by requiring that $\tilde{\bm H}_{\text{off-diag}}=0$. 
	Given that the explicit form of the canonical transformation is unknown, we expand ${\bm S}$ as a power series 
	\begin{equation}
		{\bm S}={\bm S}_1+{\bm S}_2+{\bm S}_3+\hdots\;,
	\end{equation}
	where ${\bm S}_n$ corresponds to the $n$th order of perturbation. 
	We separate the terms of different order in perturbation in $\tilde{\bm H}_{\text{off-diag}}=0$. 
	Separating the terms of different orders, we obtain algebraic equations for the matrices ${\bm S}_n$
	\begin{subequations}
		\begin{eqnarray}
			\left[{\bm H}_0,{\bm S}_1\right]&=&-{\bm H}_2\;,\label{eq:s1}\\
			\left[{\bm H}_0,{\bm S}_2\right]&=&-\left[{\bm H}_1,{\bm S}_1\right]\;,\label{eq:s2}\\
			\left[{\bm H}_0,{\bm S}_3\right]&=&-\left[{\bm H}_1,{\bm S}_2\right]-\frac{1}{3}\left[{\bm H}_2,{\bm S}_1\right]^{(2)}\;,\\
			\vdots\nonumber
		\end{eqnarray}
	\end{subequations}
	Starting with ${\bm S}_1$, we can consecutively solve for the following ${\bm S}_n$ terms. 
	And using the ${\bm S}_n$ matrices, we determine 
	\begin{eqnarray}
		\tilde{\mathcal{H}}^{\rm sw} = \tilde{\mathcal{H}}_{\text{diag}}=\sum_{j=0}^\infty \tilde{\mathcal{H}}^{(n)} \; , 
	\end{eqnarray}
	where $\tilde{\mathcal{H}}^{(n)}$ denote the $n$th order in perturbation (in this case, $K_\perp$
	
	\begin{subequations}
		\begin{eqnarray}
			\tilde{\mathcal{H}}^{(0)}&=&{\bm H}_0\\
			\tilde{\mathcal{H}}^{(1)}&=&{\bm H}_1\\
			\tilde{\mathcal{H}}^{(2)}&=&
			\left[H_2,{\bm S}_1\right]
			+\frac{1}{2}\left[{\bm H}_0,{\bm S}_1\right]^{(2)}\label{eq:h2}\\
			\tilde{\mathcal{H}}^{(3)}&=&
			\left[H_2,{\bm S}_2\right]
			+\frac{1}{2}\left[{\bm H}_1,{\bm S}_1\right]^{(2)}
			+ \frac{1}{2}\left[\left[{\bm H}_0,{\bm S}_1\right],{\bm S}_2\right]
			+\frac{1}{2}\left[\left[{\bm H}_0,{\bm S}_2\right],{\bm S}_1\right]\\
			\vdots\nonumber
		\end{eqnarray}
		\label{eq:effH_eqs}
	\end{subequations}
	
	
	Solving~(\ref{eq:s1}), we find 
	\begin{eqnarray}
		{\bm S}_1 =\frac{-K_\perp}{2(g_z h^z - 2 J_z)}\left(\begin{array}{cccccc}
			0 & 0 & 0 & 0   &  \cos\frac{{\bm \delta}_{\sf AB}  \cdot  {\bf k}}{2} &  e^{i\frac{2\pi}{3}} \cos\frac{{\bm \delta}_{\sf CA}    \cdot {\bf k}}{2} \\
			0 & 0 & 0 &  \cos\frac{{\bm \delta}_{\sf AB}  \cdot  {\bf k}}{2}   &  0 & e^{-i \frac{2\pi}{3}} \cos\frac{{\bm \delta}_{\sf BC}   \cdot  {\bf k}}{2}\\
			0 & 0 & 0 &  e^{i \frac{2\pi}{3}} \cos\frac{{\bm \delta}_{\sf CA}  \cdot  {\bf k}}{2} &  e^{-i \frac{2\pi}{3}}\cos\frac{{\bm \delta}_{\sf BC}  \cdot  {\bf k}}{2}&  0\\
			0 &  \cos\frac{{\bm \delta}_{\sf AB}  \cdot  {\bf k}}{2} &  e^{-i\frac{2\pi}{3}} \cos\frac{{\bm \delta}_{\sf CA}  \cdot  {\bf k}}{2} & 0 & 0 & 0 \\
			\cos\frac{{\bm \delta}_{\sf AB}  \cdot  {\bf k}}{2}   &  0 & e^{i \frac{2\pi}{3}} \cos\frac{{\bm \delta}_{\sf BC}  \cdot  {\bf k}}{2} & 0 & 0 & 0 \\
			e^{-i \frac{2\pi}{3}} \cos\frac{{\bm \delta}_{\sf CA}  \cdot  {\bf k}}{2} &  e^{i \frac{2\pi}{3}}\cos\frac{{\bm \delta}_{\sf BC}  \cdot  {\bf k}}{2}&  0 & 0 & 0 & 0 \\
		\end{array}
		\right)\;.
	\end{eqnarray}
	Using Eq.~(\ref{eq:s2}), we can get the second-order correction that includes the effect of the bond--symmetric exchange anisotropy
	\begin{eqnarray}
		{\bm H}^{(2)}  = \left(\begin{array}{cc}
			{\bm M}^{(2)}_{\bf k} & {\bm 0} \\
			{\bm 0} &  -{\bm M}^{(2)*}_{-{\bf k}} 
		\end{array}
		\right)
		\label{eq:Heff}
	\end{eqnarray}
	where
	\begin{eqnarray}
		{\bm M}^{(2)}_{\bf k}=-\frac{K_\perp^2}{2(g_zh^z -2J_z)}\left(\begin{array}{ccc}
			1 +\frac{\cos({\bm \delta}_{\sf CA}  \cdot  {\bf k})+\cos({\bm \delta}_{\sf AB}  \cdot  {\bf k})}{2}  & e^{-i\frac{2\pi}{3}} \cos\frac{{\bm \delta}_{\sf BC}  \cdot  {\bf k}}{2} \cos\frac{{\bm \delta}_{\sf CA}  \cdot  {\bf k}}{2} & e^{i\frac{2\pi}{3}} \cos\frac{{\bm \delta}_{\sf AB}  \cdot  {\bf k}}{2} \cos\frac{{\bm \delta}_{\sf BC}  \cdot  {\bf k}}{2}    \\
			e^{i\frac{2\pi}{3}} \cos\frac{{\bm \delta}_{\sf BC}  \cdot  {\bf k}}{2} \cos\frac{{\bm \delta}_{\sf CA}  \cdot  {\bf k}}{2}    & 1 +\frac{\cos({\bm \delta}_{\sf AB}  \cdot  {\bf k})+\cos({\bm \delta}_{\sf BC}  \cdot  {\bf k})}{2}    & e^{-i\frac{2\pi}{3}} \cos\frac{{\bm \delta}_{\sf AB}  \cdot  {\bf k}}{2} \cos\frac{{\bm \delta}_{\sf CA}  \cdot  {\bf k}}{2} \\
			e^{-i\frac{2\pi}{3}} \cos\frac{{\bm \delta}_{\sf AB}  \cdot  {\bf k}}{2} \cos\frac{{\bm \delta}_{\sf BC}  \cdot  {\bf k}}{2}  & e^{i\frac{2\pi}{3}} \cos\frac{{\bm \delta}_{\sf AB}  \cdot  {\bf k}}{2} \cos\frac{{\bm \delta}_{\sf CA}  \cdot  {\bf k}}{2}  &   1 +\frac{\cos({\bm \delta}_{\sf BC}  \cdot  {\bf k})+\cos({\bm \delta}_{\sf CA}  \cdot  {\bf k})}{2}
		\end{array}
		\right) \;.
	\end{eqnarray}
	
	
	This order of perturbation theory is sufficient to capture the effect of $K_\perp$ on band topology, 
	and we can now work exclusively with the block--diagonal Hamiltonian 
	\begin{eqnarray}
		\tilde{\bm H}_{\sf LSW} 
		\approx
		\left(\begin{array}{cc}
			{\bm M}_{\bf k} + {\bm M}^{(2)}_{\bf k} & {\bm 0} \\
			{\bm 0}  &  -{\bm M}^*_{-{\bf k}} - {\bm M}^{(2)}_{\bf k}
		\end{array}\right) \; .
		\label{eq:H.LSW.block.diagonal}
	\end{eqnarray}
	In the limit $K_\perp \ll 2\Delta$ [Eq.~(\ref{eq:condition.on.Kperp})], this approach exactly reproduces the results of the LSW 
	calculations carried out in the presence of paring terms $K_\perp$, as described 
	in Section~\ref{sec:LSW}, and illustrated in Fig.~\ref{fig:HAF.LSW.K}.
	
	\subsection{Equation of motion in the long--wavelength limit}
	\label{sec:EoM}
	
	Given the block--diagonal Hamiltonian, Eq.~(\ref{eq:H.LSW.block.diagonal}), we can derive 
	equations of motions (EoM) for magnon operators of the same form as those given in the main text.
	We start from the Heisenberg EoM  
	\begin{eqnarray}
		- i \dot{{\bm a}}^{\dagger}_{\bf k}
		= \left[ \tilde{\mathcal{H}}_{\sf LSW} , {\bm a}^\dagger_{\bf k} \right] 
		= \tilde{{\bm M}}^{\phantom{\dagger}}_{\bf k}{\bm a}^\dagger_{\bf k} \; ,
	\end{eqnarray}
	where, to accuracy $\mathcal{O}(K_\perp^2/\Delta)$,
	\begin{eqnarray}
		\tilde{\bm M}_{\bf k} = {\bm M}_{\bf k} + {\bm M}^{(2)}_{\bf k} \; .
		\label{eq:M_tilde}
	\end{eqnarray}
	We now seek expansion of this EoM in the long--wavelength limit (${\bm q} \to  {\bm 0} \equiv \Gamma $), 
	following the approach described in \cite{Yan2018}, and outlined in the main text.
	By so doing, we arrive at a theory of the (avoided) quadratic band touching between the two 
	bands which meet in the zone center.
	
	
	We start by decoupling Eq.~(\ref{eq:M_tilde}) according to the irreducible 
	representations 
	\begin{subequations}
		\begin{eqnarray}
			\Phi_{{\sf A}_1} &=&\frac{1}{\sqrt{3}}(a^\dagger_A+ a^\dagger_B+ a^\dagger_C) \;, 
		  \\
			{\bm m}_{\sf E} &=& 
			\left(\! \frac{1}{\sqrt{2}}(a^\dagger_A -a^\dagger_B),\frac{1}{\sqrt{6}}(a^\dagger_A  + a^\dagger_B  - 2 a^\dagger_C)\!\right) \;.
		\end{eqnarray} 
	\end{subequations}
	We further use a Helmholtz--Hodge decomposition to separate the two--dimensional irrep. ${\bm m}_{\sf E}$ into
	incompressible and irrotational parts 
	\begin{equation}
		{\bm m}_{\sf E} = {\bm m}_{\sf E}^\text{\sf curl} + {\bm m}_{\sf E} ^\text{\sf div}  \; .
	\end{equation}
	satisfying 
	\begin{eqnarray}
		\nabla \cdot {\bm m}_{\sf E} ^\text{\sf curl} = 0 \quad , \quad
		\nabla_\perp \cdot {\bm m}_{\sf E} ^\text{\sf div} = 0 \quad , \quad
		\nabla_\perp = (-\partial_y, \partial_x) \; .
	\end{eqnarray}
	These fields obey the coupled EoM 
	\begin{subequations}
		\begin{align}
			-i\partial_t {\bm m}_{\sf E}^\text{\sf curl}  & =  \tilde{\omega}_0 {\bm m}_{\sf E} ^\text{\sf curl} 
			- i  \left(\sqrt{3}  D_z +\frac{3 K_\perp^2}{4(g_z h^z -2 J_z)} \right) {\bm m}_{\sf E}^\text{\sf div}
			\; , \label{SMeq:EoM.incompressible}\\
			-i\partial_t {\bm m}_{\sf E}^\text{\sf div} & = - \tilde{\rho}_S \nabla (\nabla\cdot {\bm m}_{\sf E} ^\text{\sf div}) +  \tilde{\omega}_0 {\bm m}_{\sf E} ^\text{\sf div} 
			- i \left(\sqrt{3}  D_z +\frac{3 K_\perp^2}{4(g_z h^z -2 J_z)} \right){\bm m}_{\sf E}^\text{\sf curl}
			\; ,\label{SMeq:EoM.irrotational}
		\end{align}
		\label{SMeq:EoM}
	\end{subequations}
	where 
	\begin{equation}
		\tilde{\rho}_S= \frac{J_\perp}{8} \quad , \quad 
		\tilde{\omega_0}=g_z h^z-2 J_z- J_\perp -\frac{5 K_\perp^2}{4 (g_z h^z-2 J_z)}   \; . 
		\label{eq:hydrodynamic.parameters}
	\end{equation}
	
	
	It follows directly from Eq.~(\ref{SMeq:EoM}) that a mixing of incompressible and irrotational 
	excitations occurs whenever the bond--symmetric exchange anisotropy $K_\perp$, 
	or DM inteaction, $D_z$, is finite.   
	As a result, a gap opens at the ${\bf k}={\bf 0}$ point whenever either term is non--zero.
	The continuum theory predicts this gap to be
	\begin{eqnarray}
		{\Delta}_\Gamma = 2\left|\sqrt{3} D_z+\frac{3 K_\perp^2}{4(g_z h^z -2 J_z)}\right| \; .
	\end{eqnarray}
	This is consistent with the result for ${\Delta}_\Gamma$ found in LSW theory, Eq.~\ref{EQN_Gamma_point_gap}.
	
	
	It also follows from Eq.~(\ref{SMeq:EoM}) that the bond--symmetric interaction $K_\perp$ has the same 
	effect on excitations near the quadratic band touching as a positive value of DM interaction, $D^z > 0$, 
	regardless of the sign of $K_\perp$.
	This fact is consistent with the contributions to Chern numbers found in LSW theory 
	[cf. Fig.~\ref{FIG_dispersion_C_Dz_0_1} and Fig.~\ref{FIG_dispersion_C_K_1}].
	And it implies that the result for the dynamical structure factor $S({\bf q}, \omega)$ [Eq.~(11) of main text], 
	remains valid as long as appropriately--renormalised values are used for the hydrodynamic parameters 
	$\rho_S$ and $\omega_0$ [cf. Eq.~(\ref{eq:hydrodynamic.parameters})].  
	Viewed in this light, half moons make no distinction between $K_\perp$ and $D^z$.
	However the presence of magnon--pairing terms coming from $K_\perp$ would lead to a reduction 
	in the ``saturated'' moment of system, which might be observable in experiment.
	
	
	We conclude by noting that the coupled EoM describing the quadratic band--touching 
	can also be solved using Greens functions, within a Nambu (matrix) formalism.    
	In this approach pairing terms can be incorporated directly in EoM, without the need to 
	first project into the physical subspace of the model.
	

	

\end{document}